\documentclass[pdflatex,sn-mathphys-num]{sn-jnl}

\usepackage{graphicx}%
\usepackage{multirow}%
\usepackage{amsmath,amssymb,amsfonts}%
\usepackage{amsthm}%
\usepackage{mathrsfs}%
\usepackage[title]{appendix}%
\usepackage{xcolor}%
\usepackage{textcomp}%
\usepackage{manyfoot}%
\usepackage{booktabs}%
\usepackage{algorithm}%
\usepackage{algorithmicx}%
\usepackage{algpseudocode}%
\usepackage{listings}%

\theoremstyle{thmstyleone}%
\theoremstyle{thmstyletwo}%

\theoremstyle{thmstylethree}%

\usepackage{subfigure}
\begin{document}
	
	\title[Static, Cylindrically Symmetric Spacetimes in the Coincident $f(Q)$ Gravity]{Static, Cylindrically Symmetric Spacetimes in the Coincident $f(Q)$ Gravity}
	
	
	\author*[1]{\fnm{Pınar} \sur{Kirezli}}\email{pkirezli@nku.edu.tr}
	
	\author[2]{\fnm{Nilhan} \sur{Özceylan}}\email{nilhan234@yahoo.com}
	
	
	\affil*[1]{\orgdiv{Science and Art Faculty, Department of Physics}, \orgname{Tekirdağ Namık Kemal University}, \orgaddress{ \city{Tekirdağ}, \postcode{59030}, \country{Türkiye}}}
	
	\affil[2]{\orgdiv{School of Gratuated Studies, Department of Physics}, \orgname{Tekirdağ Namık Kemal University}, \orgaddress{ \city{Tekirdağ}, \postcode{59030}, \country{Türkiye}}}
	
	
	
	\abstract{In this paper we consider a static, cylindrically symmetric spacetime with coincident $f(Q)$ gravity. Since the field equation of this spacetime in symmetric teleparallel gravity is suitable for choosing the function of $f(Q)$ in the form of power series and exponential forms, perfect fluid solutions of these forms are discussed. Energy densities, directional pressures and energy conditions are plotted and analysed for a few different metric potentials. Although cosmic strings violate all energy conditions in both $f(Q)$ functions, the Levi-Civita solution violates all energy conditions in the power law function of $f(Q)$, but for exponential $f(Q)$ gravity they are satisfied in small regions.}

	
	
	
	\maketitle
	
\section{Introduction}\label{}
Einstein's general theory of relativity (GR) is a fundamental theory that explains the fourth force of nature in terms of the curvature of spacetime. GR can be constructed using a Riemannian manifold, which is a special case of the more general metric-affine geometry. In general, metric-affine geometry is described by torsion and non-metricity addition to curvature. Three of these concepts are called the Geometrical Trinity of Gravity \cite{BeltranJimenez:2019esp} and they are equivalent to each other under such circumstances as; it is GR when torsion and non-metricity vanish, teleparallel equivalent to GR (TEGR) \cite{Krssak:2018ywd} when curvature and non-metricity vanish and symmetric teleparallel equivalent to GR (STEGR) \cite{BeltranJimenez:2017tkd,Nester:1998mp} when curvature and torsion vanish. 

Although GR can perfectly explain our solar system (with Mercury's perihelion), the bending of light by the Sun, or predict black holes (which has recently been proven), it faces several problems in explaining the evolution of the universe. More recently, Supernova observations \cite{SupernovaSearchTeam:2001qse}, large-scale structure \cite{SDSS:2003eyi}, the Cosmic Microwave Background \cite{Larson:2010gs,WMAP:2010qai}, Baryonic Acoustic Oscillations \cite{SDSS:2005xqv,SDSS:2009ocz}, and the WMAP experiment \cite{WMAP:2003elm,WMAP:2003ivt}  all point to an accelerating behaviour of the Universe. Standard cosmology known as $\Lambda$-Cold-Dark-Matter ($\Lambda$-CDM) explains the accelerating expansion of the universe with dark energy which has a negative pressure. However, because the dark nature of the universe is a great mystery, researchers tend to modify GR to explain it. Thus, in constructing the new GR, the first step was to modify it by redefining the action. $f(R)$ gravity \cite{Nojiri:2003ft,Nojiri:2007as} is the best known modified theory which takes the action as a function of $R$. More recently, because of STEGR, $f(Q)$ theory \cite{BeltranJimenez:2017tkd} is also being constructed to understand our universe as well. In this way, the theory is revisited in cosmology \cite{BeltranJimenez:2019tme,Paliathanasis:2023ngs,Paliathanasis:2023nkb,Atayde:2021pgb,Esposito:2021ect,Koussour:2022irr,Koussour:2022jss,Koussour:2022wbi,Dixit:2022vyz,Sarmah:2023oum,Pradhan:2022dml,Bhar:2023xku,Bhar:2023yrf,Dimakis:2022rkd,Dimakis:2022wkj,Capozziello:2022tvv}, in black holes \cite{DAmbrosio:2021zpm,Javed:2023qve,Javed:2023vmb,Junior:2023qaq,Gogoi:2023kjt,Bahamonde:2022esv}, and in wormholes \cite{Banerjee:2021mqk,Kiroriwal:2023nul,Mustafa:2023kqt,Godani:2023nep,Mishra:2023bfe,Hassan:2022ibc,Hassan:2022hcb,Parsaei:2022wnu,Sokoliuk:2022efj,Jan:2023djj}. In addition, exact solutions of spherically symmetric spacetime in $f(Q)$ theory are analysed in \cite{Wang:2021zaz,Lin:2021uqa,Calza:2022mwt} and a fruitful review of the theory is prepared by Heisenberg in \cite{Heisenberg:2023lru}.

In addition to spherically symmetric solutions of GR, static, cylindrically symmetric metrics, which offer different possibilities to obtain some realistic results, are often studied. The pioneer of the cylindrically symmetric solution of GR is Levi-Civita (LC) \cite{Levi-Civita:1917pgo} which contains two independent parameters and was obtained very soon after Schwarzschild. Another important solution of cylindrically symmetric spacetime in GR is cosmic strings which are well known topological defects \cite{Hiscock:1985uc,Linet:1986sr,Tian:1986zz,Zofka:2007bi}. They are both exterior solution of a cylindrically symmetric source. Although, this spacetime is analysed for various modified theories such as; in Brans-Dicke \cite{Delice:2006uz}, in $f(R)$-gravity \cite{Azadi:2008qu}, in $f(G)$-gravity \cite{Houndjo:2013us} , in $f(R,G)$-gravity \cite{Zia:2019ddd}, in $f(T)$-gravity \cite{Houndjo:2012sz}, in mimetic gravity \cite{Momeni:2015aea}, in $f(R,\phi,X)$-gravity \cite{Malik:2022cyz}, and in the Einstein-Aether theory \cite{Chan:2021ivp}, it has not been studied in $f(Q)$-gravity, yet. This motivates us to analyse static, cylindrically symmetric metric in the coincident $f(Q)$ gravity.

The paper is organised as, in Section \ref{section2}, STEGR and $f(Q)$ gravity are reviewed to establish the background of our calculations. Static, cylindrically symmetric metric with perfect fluid energy-momentum tensor is defined and energy density and $r,z,\phi$ directional pressures are calculated from the field equations of $f(Q)$ gravity in Section \ref{section3}. Energy conditions, energy densities, directional pressures for different three cases are calculated and plotted for power-law and exponential form of the $f(Q)$-gravity in Section \ref{section4} and Section \ref{section5}, respectively. Our results are discussed and summarised in Section \ref{conc}.
	
\section{$f(Q)$ Gravity}\label{section2}
Against Einstein GR, more general connection called affine connection with antisymmetric part and relaxation of metricity can be defined;
\begin{eqnarray}\label{affine1}
	\Gamma^{\alpha}_{~\mu \nu}= \mathring{\Gamma}^{\alpha}_{~\mu \nu}+K^{\alpha}_{~\mu \nu}+L^{\alpha}_{~\mu \nu}
\end{eqnarray}
where $\mathring{\Gamma}^{\alpha}_{~\mu \nu}$ is Christoffel symbol, $K^{\alpha}_{~\mu \nu}$ is contorsion tensor and $L^{\alpha}_{~\mu \nu}$ is the disformation tensor. These quantities are described using the metric and affine connection as follows;
\begin{eqnarray}
	\mathring{\Gamma}^{\alpha}_{~\mu \nu}&=&\frac{1}{2}g^{\alpha \beta}\left(\partial_{\nu}g_{\mu \beta}+\partial_{\mu}g_{\nu \beta}-\partial_{\beta}g_{\mu \nu}\right),\\
	K^{\alpha}_{~\mu \nu}&=&\frac{1}{2}T^{\alpha}_{~\mu \nu}+T_{(\mu~~\nu)}^{~~\alpha},\\
	L^{\alpha}_{~\mu \nu}&=&\frac{1}{2}Q^{\alpha}_{~\mu \nu}-Q^{~~\alpha}_{(\mu ~~\nu)}
\end{eqnarray}
in which the torsion is $T^{\alpha}_{~\mu \nu}=2 \Gamma^{\alpha}_{~[\mu \nu]}$ and the non-metricity tensor is the covariant derivative with respect to affine connection;
\begin{eqnarray}\label{nonmetricity}
	Q_{\alpha\mu \nu}=\nabla_{\alpha}g_{\mu \nu}=\partial_{\alpha}g_{\mu \nu}-\Gamma^{\beta}_{~\alpha\mu}g_{\beta\nu}-\Gamma^{\beta}_{~\alpha\nu}g_{\beta\mu}.
\end{eqnarray}
Additionally, superpotential can be defined for simplicity with non-metricity vectors $Q_{\alpha}=Q_{\alpha~\mu}^{~\mu}$ and $\tilde{Q}_{\alpha}=Q^{\mu}_{~\alpha\mu}$;
\begin{eqnarray}\label{superpotential}
	P^{\alpha}_{~\mu \nu}=-\frac{1}{2}L^{\alpha}_{~\mu \nu}-\frac{1}{4}\left[g_{\mu \nu}\left(\tilde{Q}^{\alpha}-Q^{\alpha}\right)+\delta^{\alpha}_{~(\mu}Q_{\nu)}\right]
\end{eqnarray}
and the non-metricity scalar becomes more convenient form with the contraction of superpotential and non-metricity tensor;
\begin{eqnarray}
	Q=-Q_{\alpha\mu\nu}P^{\alpha\mu\nu}.
\end{eqnarray}
By the definition of Riemann tensor in GR, it will be written by the affine connection as;
\begin{eqnarray}
	R^{\alpha}_{~\beta \mu \nu}&=&\partial_{\mu}\Gamma^{\alpha}_{~\nu \beta}-\partial_{\nu}\Gamma^{\alpha}_{~\mu\beta}+\Gamma^{\alpha}_{~\mu \gamma}\Gamma^{\gamma}_{~\nu\beta}-\Gamma^{\alpha}_{~\nu \gamma}\Gamma^{\gamma}_{~\mu\beta}\\
	&=&\mathring{R}^{\alpha}_{~\beta\mu \nu}+\mathring{\nabla}_{\mu}H^{\alpha}_{~\nu \beta}-\mathring{\nabla}_{\nu}H^{\alpha}_{~\mu \beta}+H^{\alpha}_{~\mu \gamma}H^{\gamma}_{~\nu \beta}-H^{\alpha}_{~\nu \gamma}H^{\gamma}_{~\mu \beta}
\end{eqnarray}
where $H^{\alpha}_{~\mu \nu}=K^{\alpha}_{~\mu \nu}+L^{\alpha}_{~\mu \nu}$ and $\mathring{R}^{\alpha}_{~\beta\mu \nu}$ is corresponding Riemann tensor of the GR and $\mathring{\nabla}_{\mu}$ covariant derivative with respect to Christoffel symbol. If the torsion ($T^{\alpha}_{~\mu \nu}$) of the spacetime vanish the Ricci scalar becomes;
\begin{eqnarray}
	R=\mathring{R}-Q+\mathring{\nabla}_{\alpha}\left(Q^{\alpha}-\tilde{Q}^{\alpha}\right)
\end{eqnarray} 
where $\mathring{R}$ is corresponding Ricci scalar of the GR. According to this result, when total curvature of the spacetime is set zero ($R=0$), Ricci scalar of the GR becomes equivalent to non-metricity scalar plus boundary terms. This boundary term has no effect in the equation of motion and we can conclude that $\mathring{R}=Q$ which is called symmetric equivalent to general relativity (STEGR).

When constructing $f(Q) $ theory, the affine connection (\ref{affine1}) can be written in a coordinate system in which it vanishes;
\begin{eqnarray}
	\Gamma^{\alpha}_{~\mu \nu}=\frac{\partial x^{\alpha}}{\partial \zeta^{\beta}}\partial_{\mu} \partial_{\nu}\zeta^{\beta}
\end{eqnarray}
which is called coincident gauge \cite{BeltranJimenez:2017tkd} and the covariant derivative for this coordinates becomes partial derivative. The non-metricity tensor becomes with coincident gauge as;
\begin{eqnarray}
	Q_{\alpha\mu \nu}=\nabla_{\alpha}g_{\mu \nu}=\partial_{\alpha}g_{\mu \nu}.
\end{eqnarray} 
In general, the $f(Q)$-gravity is described by the action;
\begin{eqnarray}
	S=\int dx^4\sqrt{-g}\left(\frac{1}{2\kappa^2}f(Q)+L_m\right)
\end{eqnarray}
in which $f(Q)$ is any function of non-metricity and $L_m$ is the Lagrangian density of the matter. One gets variation of the action with respect to metric and affine connection \cite{Adak:2005cd,Adak:2008gd}
\begin{eqnarray}
	\frac{2}{\sqrt{-g}}\nabla_{\alpha}\left(\sqrt{-g}f_QP^{\alpha}_{~\mu \nu}\right)+\frac{1}{2}fg_{\mu \nu}+f_Q\left(P_{\mu \alpha\beta}Q_{\nu}^{~\alpha\beta}-2Q_{\alpha\beta\mu}O^{\alpha\beta}_{~~\nu}\right)&=&T_{\mu \nu}\\
	\nabla_{\mu}\nabla_{\nu}\left(\sqrt{-g}f_QP^{\mu \nu}_{~~\alpha}\right)&=&0
\end{eqnarray}
where $f_Q=\frac{\partial f}{\partial Q}$ and the stress-energy-momentum tensor $T_{\mu \nu}=-\frac{2}{\sqrt{-g}}\frac{\delta\sqrt{-g}L_m}{\delta\sqrt{g_{\mu \nu}}}$. These equations can be written in more convenient form as \cite{Lin:2021uqa};
\begin{eqnarray}\label{fieldeqn}
	G_{\mu \nu}=f_Q\mathring{G}_{\mu \nu}+\frac{1}{2}g_{\mu \nu}\left(f-f_Q Q\right)+2f_{QQ}P^{\alpha}_{~\mu \nu}\partial_{\alpha}Q=T_{\mu \nu}
\end{eqnarray}
in which $f_{QQ}=\frac{\partial}{\partial Q}\frac{\partial f}{\partial Q}$ and $\mathring{G}_{\mu\nu}$ is Einstein tensor in GR. It obviously seems that $f(Q)=Q$ the above equation reduces to GR.

\section{Static, cylindrically symmetric metric and field equations in $f(Q)$ gravity }\label{section3}
In general, static, cylindrically symmetric spacetime can be examined as;
\begin{eqnarray}\label{metric}
	ds^2=-e^{2U(r)}dt^2+e^{2\left(K(r)-U(r)\right)}\left(dr^2+dz^2\right)+W(r)^2e^{-2U(r)}d\phi^2
\end{eqnarray}
where $U,K,W$ are the function of $r$ and it is usually assumed that $t\in (-\infty,\infty)$, $r\in [0,\infty)$, $z\in (-\infty,\infty) $, $\phi \in [0,2\pi)$.

While defining metric, we will introduce the components of symmetric teleparallel tensor. Non-zero component of the non-metricity tensor (\ref{nonmetricity}) are obtained;
\begin{eqnarray}
	&&Q_{rtt}=-2e^{2U}U',~~~~~~~Q_{rrr}=Q_{rzz}=2e^{2(K-U)}\left(K'-U'\right)\nonumber\\
	&&Q_{r\phi\phi}=2We^{-2U}\left(WU'-W'\right)
\end{eqnarray}
where $(')$ indicates the derivation with respect to coordinate $r$. Non null components of the disformation tensor read;
\begin{eqnarray}
	&&	L^{t}_{~rt}=-U',~~~~L^{r}_{~tt}=e^{4U-2K}U',~~~~L^{r}_{~rr}=L^{r}_{~rz}=-K'+U'~~~~	L^{r}_{~zz}=K'-U',\nonumber\\
	&&L^{r}_{~\phi\phi}=e^{-2K}W\left(WU'-W'\right),~~~~L^{\phi}_{~r\phi}=\frac{1}{W}\left(WU'-W'\right)
\end{eqnarray}
and non-zero component of (\ref{superpotential}) are;
\begin{eqnarray}
	&&	P^{t}_{~rt}=\frac{4WU'-2WK'-W'}{4W},~~~~P^{r}_{~tt}=\frac{e^{4U-2K}}{2W}\left(WK'-2WU'+W'\right),\nonumber\\
	&&	P^{r}_{~rz}=\frac{-W'}{4W},~~~~P^{r}_{\phi\phi}=\frac{e^{-K}W^2K'}{2},~~~~P^{\phi}_{~r\phi}=\frac{-2WK'+W'}{4W},~~~~P^{r}_{~zz}=\frac{W'}{2W}.~~~~~~
\end{eqnarray}
Finally we get non-metricity scalar as;
\begin{eqnarray}
	Q=\frac{2e^{2(U-K)}\left(W'K'-WU'^2\right)}{W}
\end{eqnarray}
and from equation (\ref{fieldeqn}), the field equations are obtained;
\begin{eqnarray}
	G_{tt}&=&\frac{1}{2 W}\Bigg[\bigg(-2 Q' \left(K' W-2 U' W+W'\right) f_{QQ}e^{-2 K+4 U}+e^{2 U} W \left(Q f_{Q}-f\right)\nonumber\\
	&&+4 \left(-\frac{U'^{2} W}{2}+U' W'-\frac{W K''}{2}+U'' W-\frac{W''}{2}\right) f_{Q}\bigg)  \Bigg]=T_{tt}~~~~~~\\
	G_{rr}&=&-\frac{1}{2W}\left(We^{2(K-U)}\left(Qf_Q-f\right)+2f_Q\left(U'^2W-W'K'\right)\right)=T_{rr}\\
	G_{zz}&=&\frac{1}{2W}\left(-We^{2(K-U)}\left(Qf_Q-f\right)+2f_Q\left(W''+WU'^2-W'K'\right)+2Q'f_{QQ}W'\right)\nonumber\\
	&=&T_{zz}\\
	G_{\phi\phi}&=&\frac{W^2}{2}\left(-e^{-2U}\left(Qf_Q-f\right)+2e^{-2K}\left(U'^2f_Q+Q'K'f_{QQ}+K''f_Q\right)\right)=T_{\phi\phi}.
\end{eqnarray}  
Unlike the field equations of spherically symmetric spacetimes, the field equations of cylindrically symmetric spacetimes do not contain cross terms like $G_{r\theta}$, which restricts the function of $f$ to linear form. 
Stress-energy tensor of a perfect fluid is defined as;
\begin{eqnarray}
	T_{\mu \nu}=\left(\rho+p_n\right)u_{\mu}u_{\nu}+p_ng_{\mu \nu}
\end{eqnarray}
where $u^{\mu}$ is a four velocity and satisfies $u^{\mu}u_{\mu}=-1$, and $n$ denotes directional pressure of the coordinates $r,z,\phi$. Energy density $\rho$, radial pressure, axial pressure and azimuthal pressure are obtained as;
\begin{eqnarray}
	\rho&=&	\frac{1}{2 W^3}\Bigg(4 f_{QQ} e^{4 U-4 K} \left(W K'-2 W U'+W'\right) \bigg(W K'' W'+2 W K' U' W'\nonumber\\
	&&+2 W^2 K' U'^2+W K' W''-K' W'^2-2 W K'^2 W'-2 W^2 U'^3-2 W^2 U' U''\bigg)\nonumber\\
	&&-W^2 \bigg(2 f_{Q} e^{2 U-2 K} \left(W K''-2 W U''-2 U' W'+W U'^2+W''\right)\nonumber\\
	&&-W \left(Q f_{Q}-f\right)\bigg)\Bigg),\label{rho}\\
	p_r&=&\frac{1}{2} \left(f_{Q} \left(\frac{e^{2 U-2 K} \left(2 K' W'-2 W U'^2\right)}{W}-Q\right)+f\right),\label{pr}\\
	p_z&=&\frac{1}{2W^3} \Bigg(4 f_{QQ} W' e^{4 U-4 K} \bigg(2 W^2 U' \left(-K' U'+U''+U'^2\right)+K' W'^2\nonumber\\
	&&+W \left(W' \left(2 K' \left(K'-U'\right)-K''\right)-K' W''\right)\bigg),\label{pz}\nonumber\\
	&&+W^3f_{Q} \left(\frac{2 e^{2 U-2 K} \left(-K' W'+W U'^2+W''\right)}{W}-Q\right)+W^3f\Bigg),\\
	p_{\phi}&=&\frac{1}{2W^2} \Bigg(4 f_{QQ} K' e^{4 U-4 K} \bigg(2 W^2 U' \left(-K' U'+U''+U'^2\right)+K' W'^2\nonumber\\
	&&+W \left(W' \left(2 K' \left(K'-U'\right)-K''\right)-K' W''\right)\bigg)\nonumber\\
	&&+W^2f_{Q} \left(2 e^{2 U-2 K} \left(K''+U'^2\right)-Q\right)+W^2f\Bigg). \label{pphi}
\end{eqnarray}

Furthermore, inequalities of energy conditions for a spacetime are introduced to analyse the validity of the solution, which are described as;
\begin{itemize}
	\item Null energy condition (NEC): $\rho+p_n\geq 0$,
	\item Weak energy condition (WEC): $\rho\geq 0$ with $\rho+p_n\geq 0$,
	\item Dominant energy condition (DEC): $\rho-|p_n|\geq 0$ with $\rho\geq 0$,
	\item Strong energy condition (SEC): $\rho+\sum_{n}p_n\geq 0$ with $\rho+p_n\geq 0$.
\end{itemize}
Violation of the null energy condition in the cylindrically symmetric solution indicates the existence of cylindrical wormholes \cite{Malik:2022cyz} and, in general, violation of the strong energy condition represents an accelerating expansion of the universe \cite{Mandal:2020lyq,De:2022wmj}. 

We will analyse two different forms of the function $f(Q)$ that are commonly studied i.e. $f=\alpha+\beta Q^n$ and $f(Q)=\beta Q e^{ \frac{\lambda}{Q}}$ where $\alpha, \beta, n, \lambda$ are real constants. The power-law solution is induced GR when $\beta=n=1$ and $\alpha=0$ and the GR solution with cosmological constant when $\beta=n=1$ and $\alpha=2\lambda$. The exponential form of $f(Q)$ is the GR when $\beta=1$ and $\lambda=0$.

\section{$f(Q)=\alpha+\beta Q^n$ Gravity}\label{section4}
While the function of f is chosen as $f=\alpha+\beta Q^n$, we will discuss a few known cylindrically symmetric solutions.
\subsection{Case 1}
\begin{figure}[!h]
	\centering
	\subfigure[]{
		\includegraphics[width=.4\textwidth]{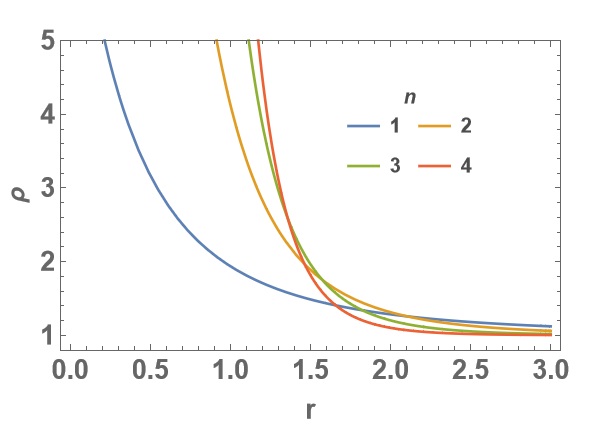}
	}
	\subfigure[]{
		\includegraphics[width=.4\textwidth]{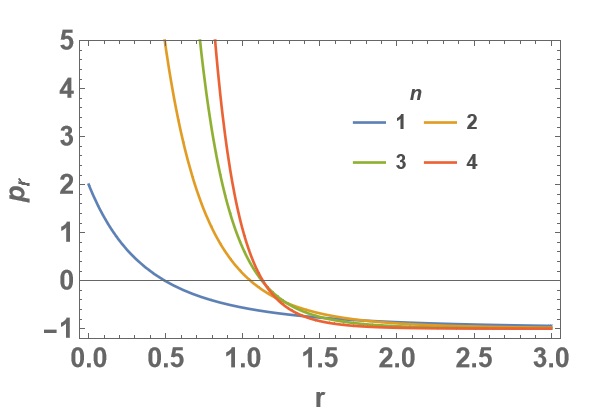}
	}
	
	\subfigure[]{
		\includegraphics[width=.4\textwidth]{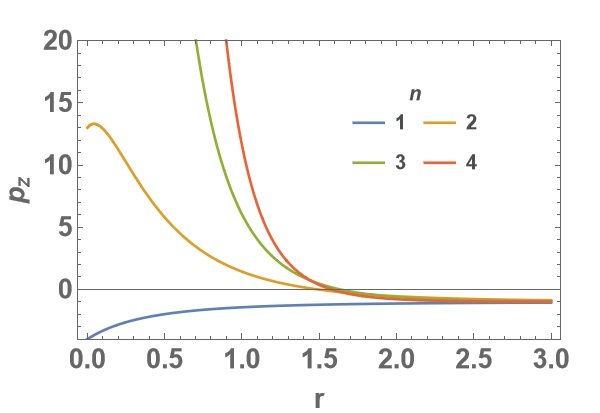}
	}
	\subfigure[]{
		\includegraphics[width=.4\textwidth]{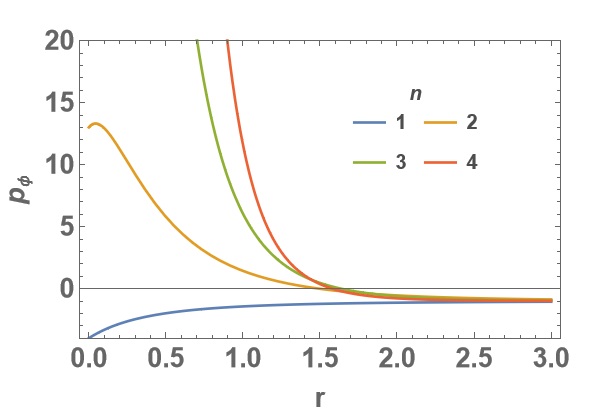}
	}
	\caption{Energy density and directional pressures in Case I for power law of function f with $W_0=W_1=U_0=1$, $\beta=-1$ and $\alpha=-2$.}
	\label{case1_1}
\end{figure}

We will start with the metric coefficients are;
\begin{eqnarray}
	W&=&W_0 r+W_1\label{wc1}\\
	K&=& \ln W\label{kc1}\\
	U&=& \ln\left(W+U_0r\right)\label{uc1}
\end{eqnarray}
in which $W_0, W_1, U_0$ are real constants. Static, cylindrically symmetric line element becomes;
\begin{eqnarray}
	ds^2=-\left(U_0r+W_0r+W_1\right)^2dt^2+\frac{\left(W_0 r+W_1\right)^2}{\left(U_0r+W_0r+W_1\right)^2}\left(dr^2+dz^2+d\phi^2\right).
\end{eqnarray}
Substituting the metric coefficients in equations (\ref{rho}-\ref{pphi}),
\begin{eqnarray}
	\rho&=&\frac{1}{2} \Bigg(-\alpha-\frac{\beta 2^n \left(\frac{U_0 W_1 (2 r W_0 (U_0+W_0)+W_1 (U_0+2 W_0))}{(r W_0+W_1)^4}\right)^n}{(2 r W_0 (U_0+W_0)+W_1 (U_0+2 W_0))^2} \bigg(2 r W_0 W_1 (U_0+W_0)\nonumber\\
	&& \big(\left(2 n^2+n+2\right) U_0+4 (n (3 n-2)+1) W_0\big)\nonumber\\
	&&+W_1^2 \left(2 \left(2 n^2+n+2\right) U_0 W_0+(2 n+1) U_0^2+4 (n (3 n-2)+1) W_0^2\right)\nonumber\\
	&&+4 (n (3 n-2)+1) r^2 W_0^2 (U_0+W_0)^2\bigg)\Bigg),\\
	p_r&=&\frac{1}{2} \left(\alpha-2^n\beta (2n-1) \left(\frac{U_0 W_1 (2 r W_0 (U_0+W_0)+W_1 (U_0+2 W_0))}{(r W_0+W_1)^4}\right)^n\right),\\
	p_z&=&\frac{1}{2} \left(\alpha+\beta 2^n \left(\frac{U_0 W_1 (2 r W_0 (U_0+W_0)+W_1 (U_0+2 W_0))}{(r W_0+W_1)^4}\right)^n\right)\nonumber\\
	&&-\frac{\beta 2^n (n-1) n U_0 W_0^2 W_1 (r (U_0+W_0)+W_1)^2 (3 r W_0 (U_0+W_0)+W_1 (U_0+3 W_0))}{(r W_0+W_1)^8}\nonumber\\
	&&\times\left(\frac{U_0 W_1 (2 r W_0 (U_0+W_0)+W_1 (U_0+2 W_0))}{(r W_0+W_1)^4}\right)^{n-2}\\
	p_{\phi}&=&\frac{1}{2} \left(\alpha+\beta 2^n \left(\frac{U_0 W_1 (2 r W_0 (U_0+W_0)+W_1 (U_0+2 W_0))}{(r W_0+W_1)^4}\right)^n\right)\nonumber\\
	&&-\frac{\beta 2^n (n-1) n U_0 W_0^2 W_1 (r (U_0+W_0)+W_1)^2 (3 r W_0 (U_0+W_0)+W_1 (U_0+3 W_0))}{(r W_0+W_1)^8}\nonumber\\
	&&\times\left(\frac{U_0 W_1 (2 r W_0 (U_0+W_0)+W_1 (U_0+2 W_0))}{(r W_0+W_1)^4}\right)^{n-2}.
\end{eqnarray}
In this case, the axial pressure and the azimuthal pressure are equal. Although the energy density $\rho$ is always positive, the radial pressure $p_r$ is initially positive but then becomes negative for large $r$ values. Axial pressure $p_z$ and azimuthal pressure $p_{\phi}$ are always negative for $n=1$, but initially positive and then become negative for other $n$ values, which are shown in Fig.\ref{case1_1}.

\begin{figure}[!h]
	\centering
	\subfigure[]{
		\includegraphics[width=.3\textwidth]{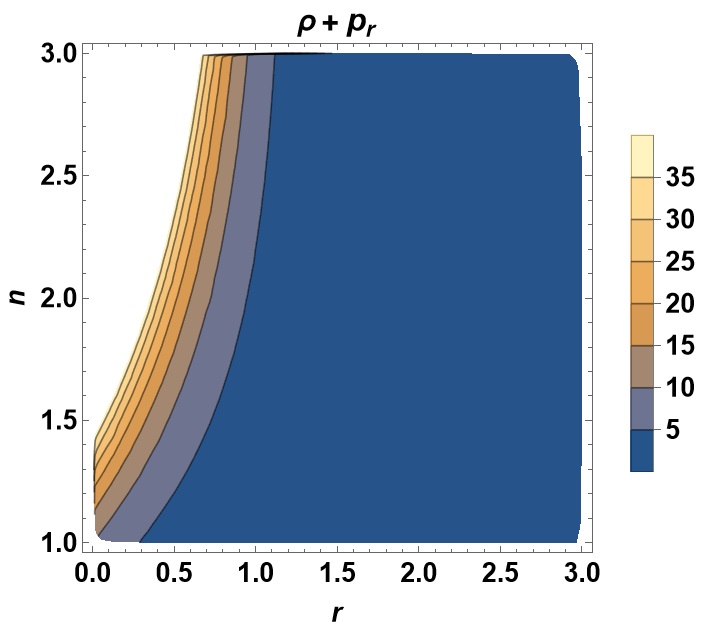}
		\label{necrc1}	}
	\subfigure[]{
		\includegraphics[width=.3\textwidth]{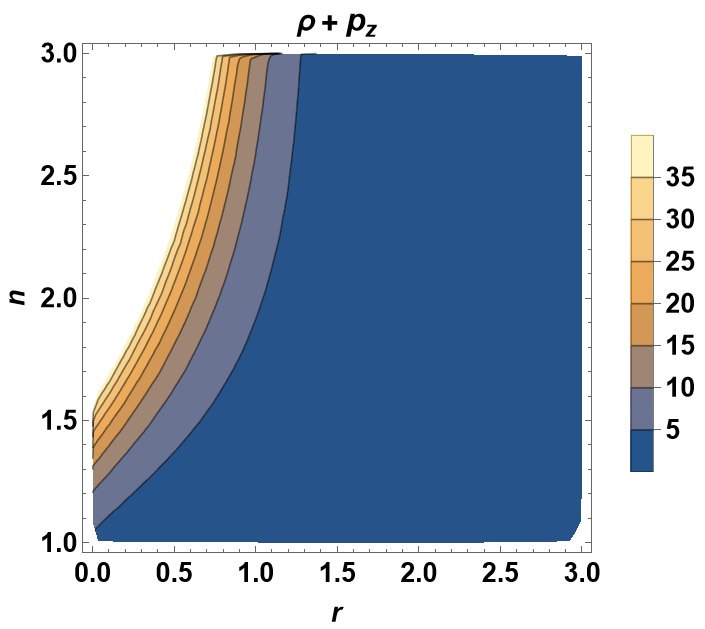}
		\label{neczc1}	}
	\subfigure[]{
		\includegraphics[width=.3\textwidth]{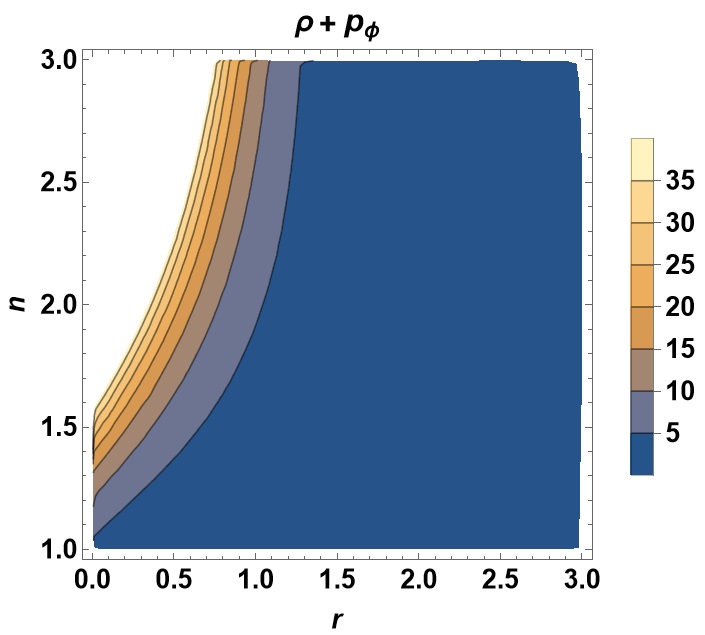}
		\label{necphic1}		}
	\caption{$\rho+p_r$, $\rho+p_z$, $\rho+p_{\phi}$ of Case I of $f=\alpha+\beta Q^{n}$ with $W_0=W_1=U_0=1$, $\beta=-1$ and $\alpha=-2$.}
	\label{case1_2}
\end{figure}

\begin{figure}[!h]
	\centering
	\subfigure[]{
		\includegraphics[width=.3\textwidth]{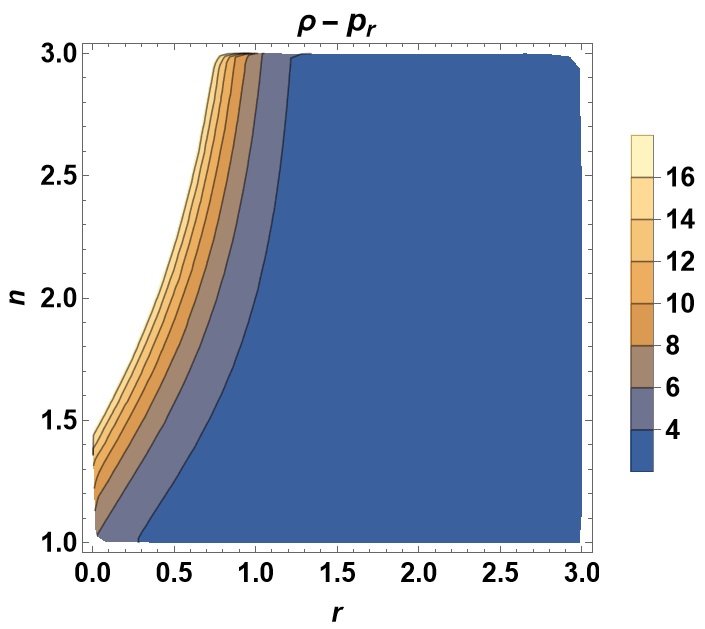}
		\label{decrc1}	}
	\subfigure[]{
		\includegraphics[width=.3\textwidth]{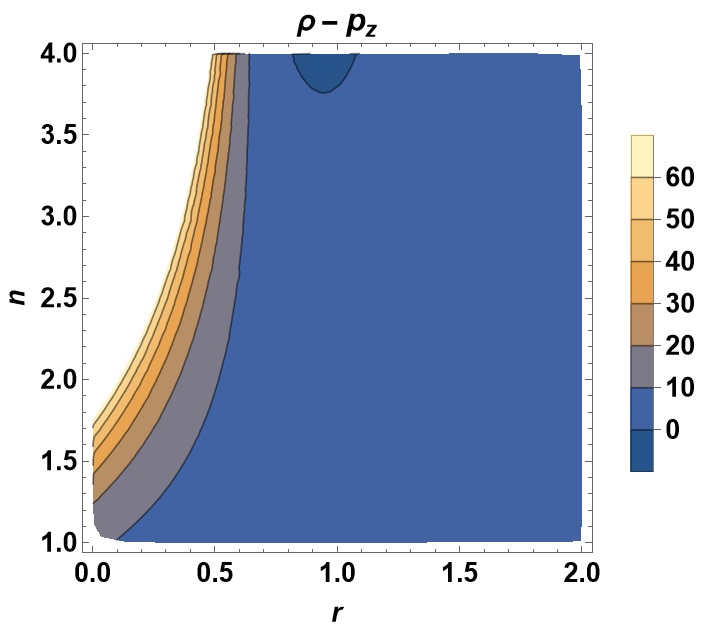}
		\label{deczc1}	}
	\subfigure[]{
		\includegraphics[width=.3\textwidth]{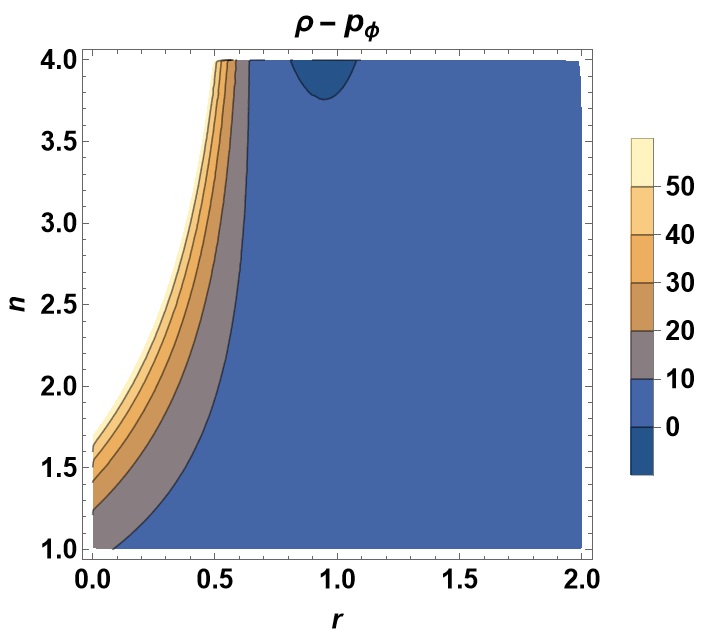}
		\label{decphic1}	}
	\caption{$\rho-p_r$, $\rho-p_z$, $\rho-p_{\phi}$ of Case 1 of $f=\alpha+\beta Q^{n}$ with $W_0=W_1=U_0=1$, $\beta=-1$ and $\alpha=-2$.}
	\label{case1_3}
\end{figure}

\begin{figure}[!h]
	\centering
	\subfigure[]{
		\includegraphics[width=.3\textwidth]{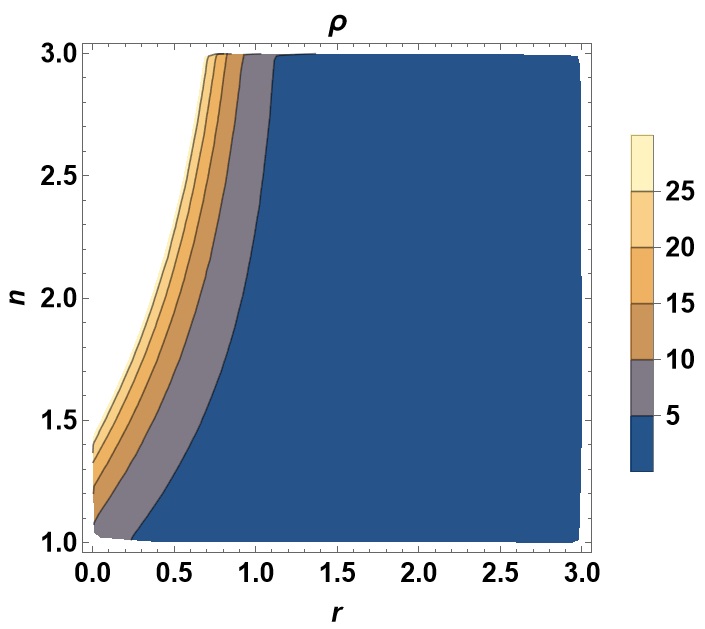}
		\label{wecc1}}
	\subfigure[]{
		\includegraphics[width=.3\textwidth]{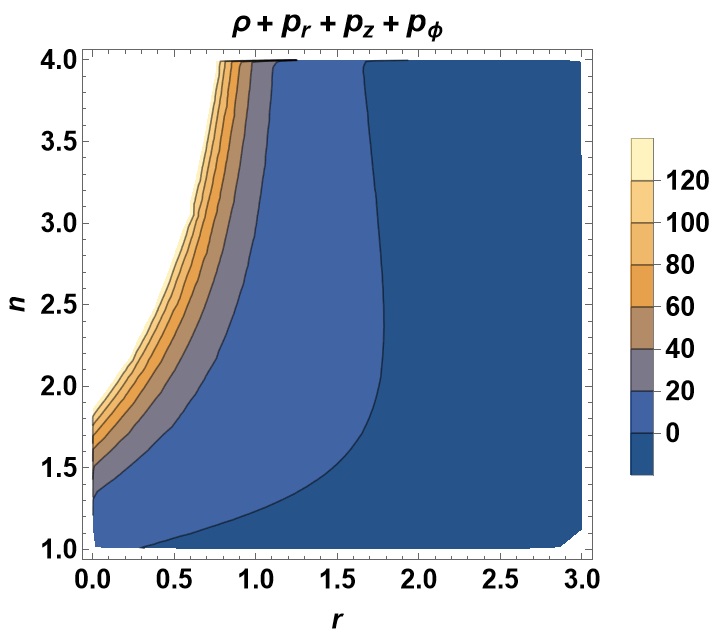}
		\label{secc1}}
	\caption{$\rho$, and $\rho+p_r+p_z+p_{\phi}$ of Case 1 of $f=\alpha+\beta Q^{n}$ with $W_0=W_1=U_0=1$, $\beta=-1$ and $\alpha=-2$.}
	\label{case1_4}
\end{figure}

Although, NEC and WEC are satisfied for all directions in Figs. \ref{case1_2} and \ref{wecc1}, DEC for radial pressure is always satisfied in Fig. \ref{decrc1}, DEC for azimuthal and axial pressure is mostly satisfied but almost $n=4$ and $r=1$ they are violated in Fig. \ref{deczc1} and Fig \ref{decphic1} which means that DEC is violated in this case. On the other hand, SEC is violated especially for $r>1.8$ for all values of $n$ in Fig. \ref{secc1}.

\subsection{Case II}
In this case we introduce metric potentials as;
\begin{eqnarray}
	W&=&r \label{wc2}\\
	K&=&K_0\ln W \label{kc2}\\
	U&=& U_0\ln W \label{uc2}
\end{eqnarray}
where $K_0$ and $U_0$ are constants and this metric is known as Levi-Civita (LC) solution in GR. The metric coefficient of $U$ is Newtonian potential of an infinitely uniform line source in GR. Energy density and directional pressures for Case II become;
\begin{eqnarray}
	\rho&=&\frac{1}{2} \Bigg(-\alpha-\frac{\beta 2^n \left(\left(U_0^2-K_0\right) r^{-2 K_0+2 U_0-2}\right)^n}{K_0-U_0^2}\big(2 K_0^2 (n-1) n-2 K_0 (n-1) n (3 U_0-2)\nonumber\\
	&&+K_0+(4 (n-1) n-1) U_0^2-6 (n-1) n U_0+2 (n-1) n\big) \Bigg),\\
	p_r&=&\frac{1}{2} \left(\alpha-\beta 2^n (2 n-1) \left(\left(U_0^2-K_0\right) r^{-2 K_0+2 U_0-2}\right)^n\right),\\
	p_z&=&\frac{1}{2} \bigg(\alpha+\frac{\beta 2^n }{K_0-U_0^2}\left(2 K_0 (n-1) n+K_0-2 (n-1) n U_0+2 (n-1) n-U_0^2\right)\nonumber\\
	&& \left(\left(U_0^2-K_0\right) r^{-2 K_0+2 U_0-2}\right)^n\bigg),~~~~\\
	p_{\phi}&=&\frac{1}{2} \bigg(\alpha+\frac{\beta 2^n }{K_0-U_0^2}\left(2 K_0^2 (n-1) n+2 K_0 n (n (-U_0)+n+U_0-1)+K_0-U_0^2\right)\nonumber\\
	&& \left(\left(U_0^2-K_0\right) r^{-2 K_0+2 U_0-2}\right)^n\bigg).~~~~~~
\end{eqnarray}

\begin{figure}[!h]
	\centering
	\subfigure[]{
		\includegraphics[width=.4\textwidth]{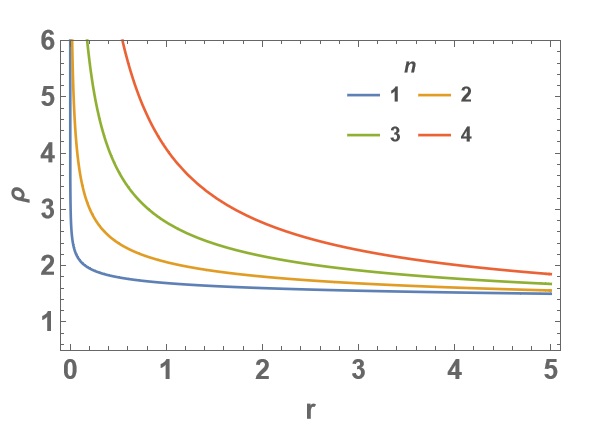}
	}
	\subfigure[]{
		\includegraphics[width=.4\textwidth]{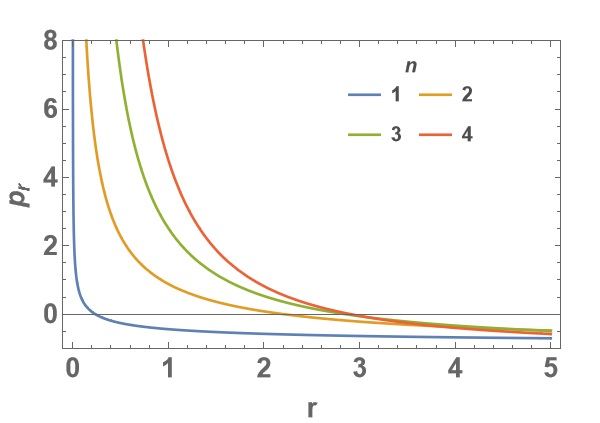}
	}
	
	\subfigure[]{
		\includegraphics[width=.4\textwidth]{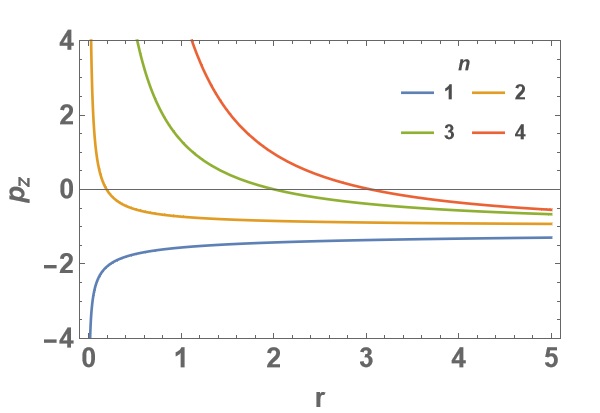}
	}
	\subfigure[]{
		\includegraphics[width=.4\textwidth]{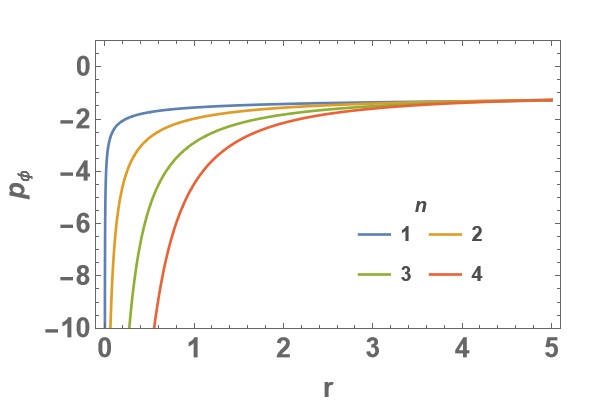}
	}
	\caption{Energy density and directional pressures in Case II for power law of function f with $\alpha=-2$, $U_0=0.4$, $K_0=-0.4$ and $\beta=-1$.}
	\label{case2_1}
\end{figure}

The energy density of the Case II is always positive for different values of $n$. Although, the radial and axial pressures are positive for small values of $r$ and then become negative, but $p_\phi$ is always negative for few different values of $n$ with chosen values of parameters which are shown in Fig. \ref{case2_1}. Furthermore, the axial pressure for $n=1$ always has negative values, unlike the other values of $n$.

\begin{figure}[!h]
	\centering
	\subfigure[]{
		\includegraphics[width=.3\textwidth]{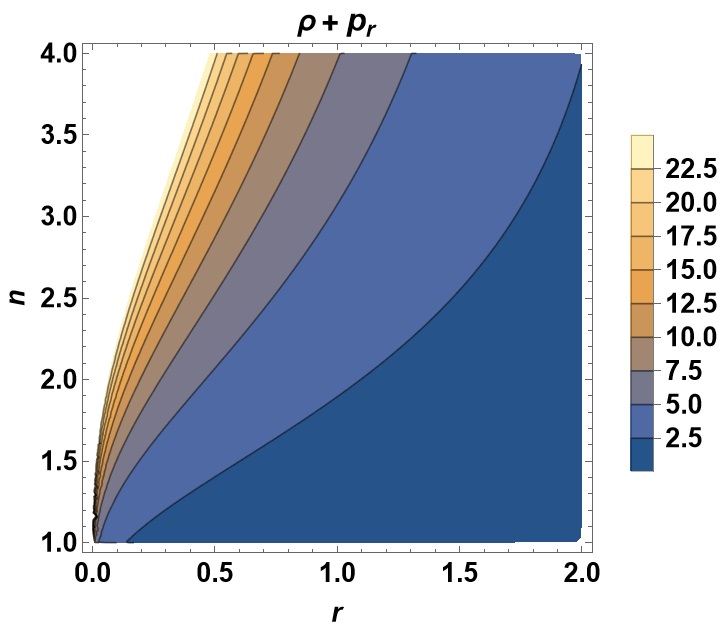}
		\label{necrc2}}
	\subfigure[]{
		\includegraphics[width=.3\textwidth]{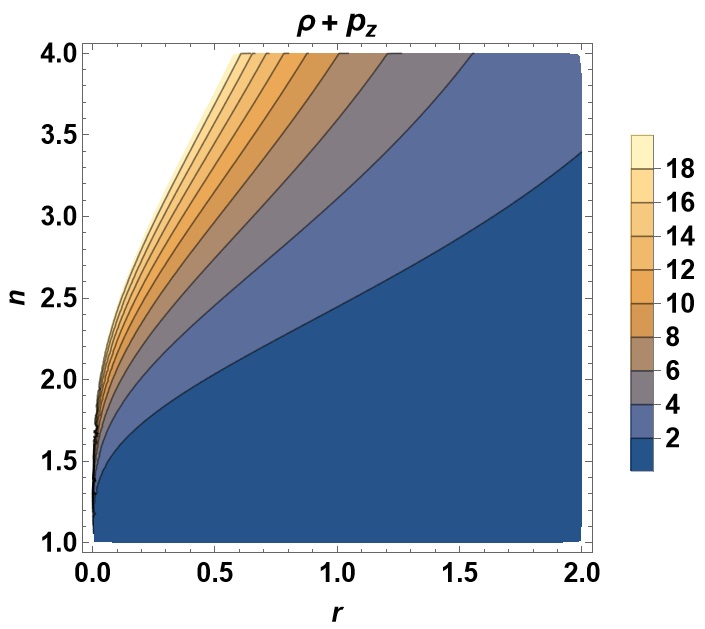}
		\label{neczc2}}
	\subfigure[]{
		\includegraphics[width=.3\textwidth]{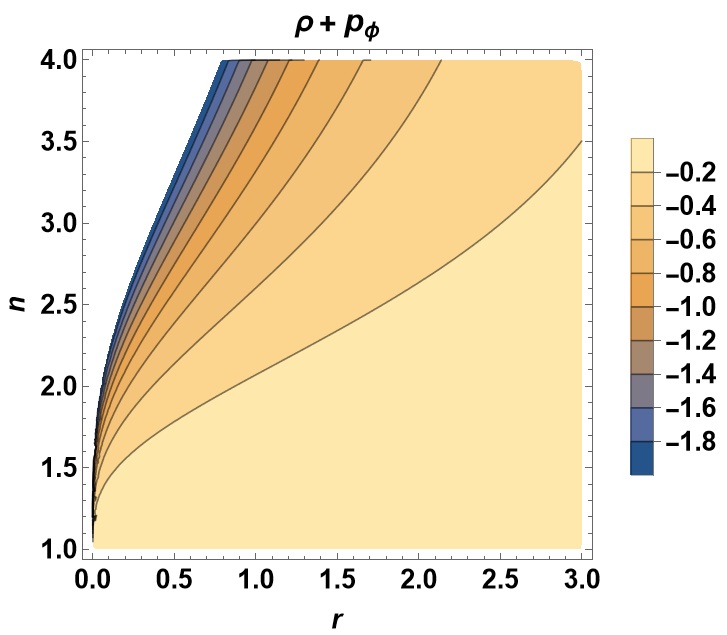}
		\label{necphic2}	}
	\caption{$\rho+p_r$, $\rho+p_z$, $\rho+p_{\phi}$ of Case II of $f=\alpha+\beta Q^{n}$ with $\alpha=-2$, $U_0=0.4$, $K_0=-0.4$ and $\beta=-1$.}
	\label{case2_2}
\end{figure}

\begin{figure}[!h]
	\centering
	\subfigure[]{
		\includegraphics[width=.3\textwidth]{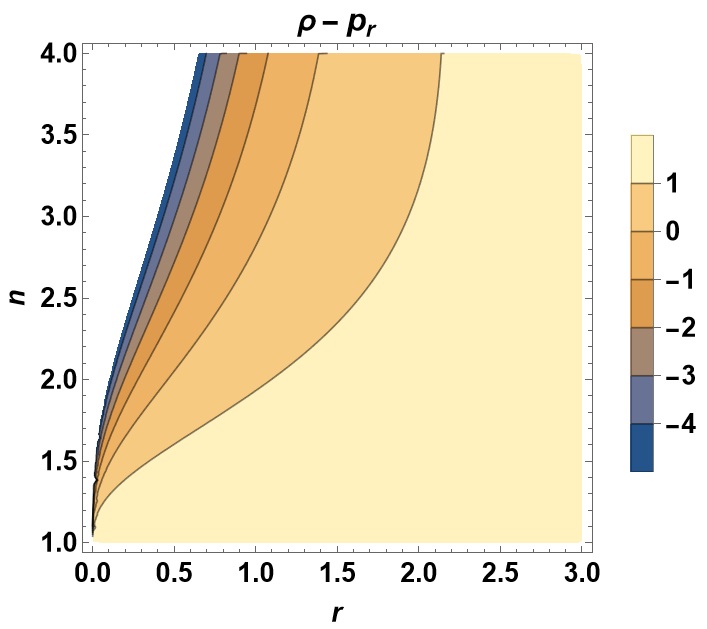}
		\label{decrc2}	}
	\subfigure[]{
		\includegraphics[width=.3\textwidth]{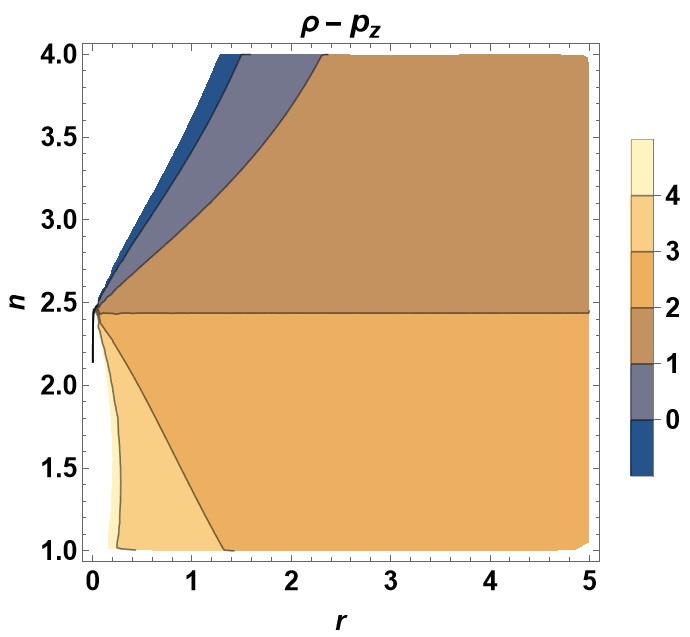}
		\label{deczc2}	}
	\subfigure[]{
		\includegraphics[width=.3\textwidth]{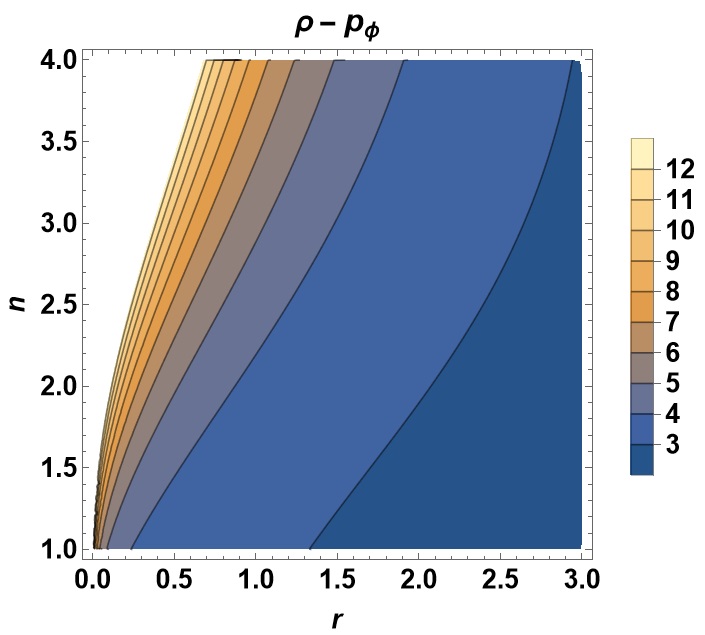}
		\label{decphic2}	}
	\caption{$\rho-p_r$, $\rho-p_z$, $\rho-p_{\phi}$ of Case II of $f=\alpha+\beta Q^{n}$ with $\alpha=-2$, $U_0=0.4$, $K_0=-0.4$ and $\beta=-1$.}
	\label{case2_3}
\end{figure}

\begin{figure}[!h]
	\centering
	\subfigure[]{
		\includegraphics[width=.3\textwidth]{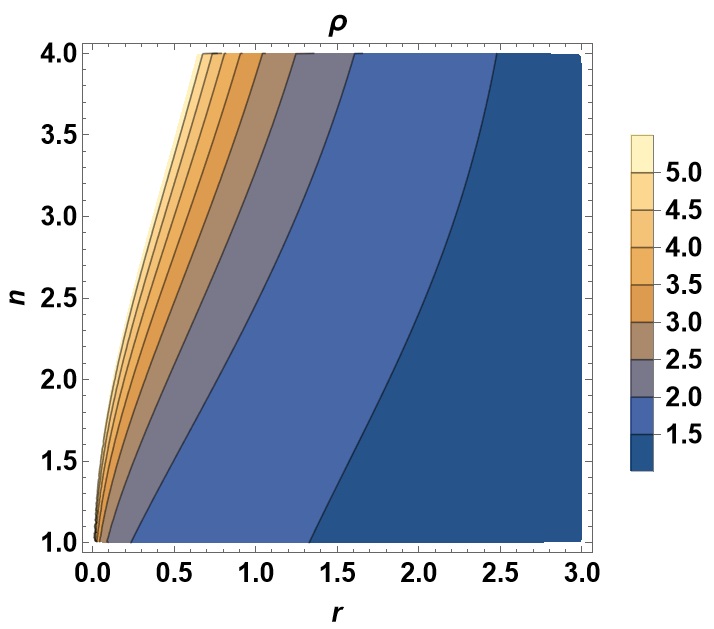}
		\label{wecc2}}
	\subfigure[]{
		\includegraphics[width=.3\textwidth]{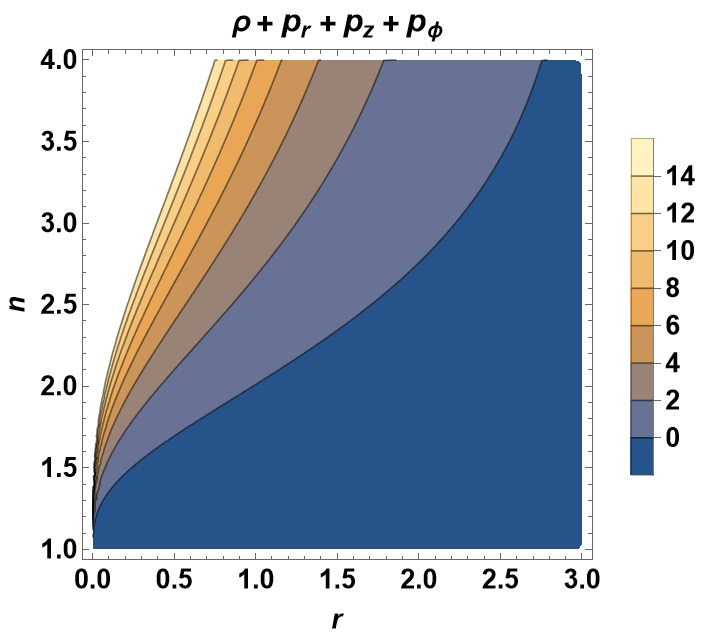}
		\label{secc2}}
	\caption{$\rho$, and $\rho+p_r+p_z+p_{\phi}$ of Case II of $f=\alpha+\beta Q^{n}$ with $\alpha=-2$, $U_0=0.4$, $K_0=-0.4$ and $\beta=-1$.}
	\label{case2_4}
\end{figure}

The obligatory statements for theenergy conditions are shown in Figs.\ref{case2_1}-\ref{case2_4}. NEC is satisfied where $\rho+p_r$ and $\rho+p_z$ are positive in Fig. \ref{necrc2} and \ref{neczc2}, whereas NEC is violated for $\phi$ direction which is shown in Fig. \ref{necphic2}. DEC is automatically violated for the azimuthal direction even though $\rho-p_{\phi} \geq 0$ in Fig. \ref{decphic2}. In particular, the DEC for other directions are satisfied, but violated for large $n$ and small values of $r$, as shown in Figs. \ref{decrc2} and \ref{deczc2}, respectively. In addition, since $\rho$ is always greater than zero in Fig. \ref{wecc2}, WEC is satisfied for the $r,z$ directions but violated for the azimuthal direction. SEC is particularly violated, but is satisfied for large values of $n$, as shown in Fig. \ref{secc2}. We can conclude that, NEC, WEC, DEC and SEC are violated in the LC solution.

\subsection{Case III}
\begin{figure}[!h]
	\centering
	\subfigure[]{
		\includegraphics[width=.4\textwidth]{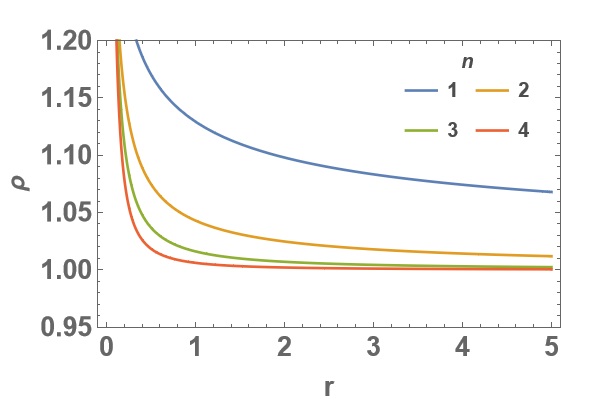}
	}
	\subfigure[]{
		\includegraphics[width=.4\textwidth]{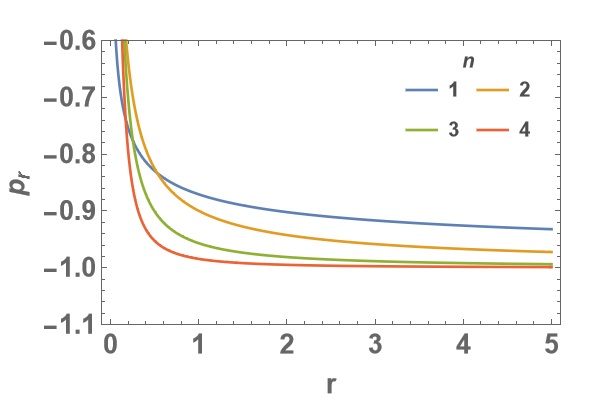}
	}
	
	\subfigure[]{
		\includegraphics[width=.4\textwidth]{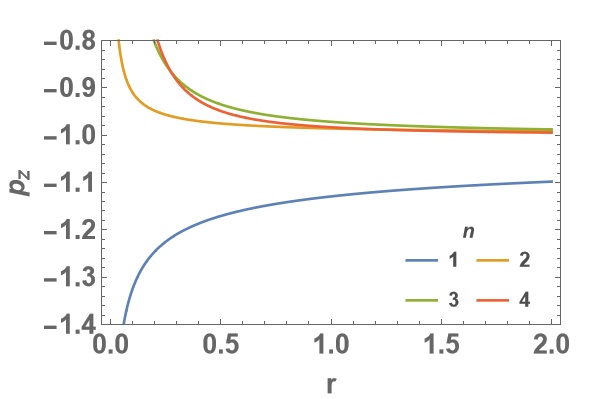}
	}
	\subfigure[]{
		\includegraphics[width=.4\textwidth]{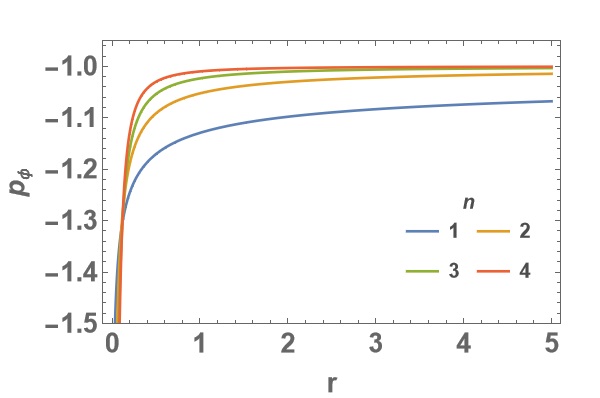}
	}
	\caption{Energy density and directional pressures in Case III for power law of function f with $\alpha=-2$, $K_0=-0.4$, $U_0=W_0=0.4$ and $\beta=-1$.}
	\label{case3_1}
\end{figure}

Another important solution in cylindrically symmetric spacetime is cosmic strings where the metric coefficients are;
\begin{eqnarray}
	W&=&W_0 r, \label{wc3}\\
	K&=&K_0 \ln W,\label{kc3}\\
	U&=&U_0 \ln W\label{uc3}
\end{eqnarray}
and line element is obtained;
\begin{eqnarray}
	ds^2=-\left(W_0 r\right)^{2U_0}dt^2+\left(W_0 r\right)^{2K_0-2U_0}\left(dr^2+dz^2\right)+W_0^2 r^2\left(W_0 r\right)^{-2U_0} d\phi^2.
\end{eqnarray}
It is obviously seem that, when $W_0=1$ the metric reduces LC solution. Energy density and directional pressures of the cosmic string become;
\begin{eqnarray}
	\rho&=& \frac{1}{2} \Bigg(-\alpha-\frac{\beta 2^n\left(\frac{\left(U_0^2-K_0\right) (r W_0)^{2 U_0-2 K_0}}{r^2}\right)^n}{K_0-U_0^2} \Big(2 K_0^2 (n-1) n-2 K_0 (n-1) n (3 U_0-2)\nonumber\\
	&&+K_0+(4 (n-1) n-1) U_0^2-6 (n-1) n U_0+2 (n-1) n\Big) \Bigg) ~~~~~~\\
	p_r&=&\frac{1}{2} \left(\alpha-\beta 2^n (2 n-1) \left(\frac{\left(U_0^2-K_0\right) (r W_0)^{2 U_0-2 K_0}}{r^2}\right)^n\right)\\
	p_z&=&\frac{1}{2} \Bigg(\alpha+\frac{\beta 2^n }{K_0-U_0^2}\left(2 K_0 (n-1) n+K_0-2 (n-1) n U_0+2 (n-1) n-U_0^2\right)\nonumber\\
	&& \left(\frac{\left(U_0^2-K_0\right) (r W_0)^{2 U_0-2 K_0}}{r^2}\right)^n\Bigg)\\
	p_{\phi}&=&\frac{1}{2} \bigg(\alpha+\frac{\beta 2^n }{K_0-U_0^2}\left(2 K_0^2 (n-1) n+2 K_0 n (n (-U_0)+n+U_0-1)+K_0-U_0^2\right)\nonumber\\
	&& \left(\frac{\left(U_0^2-K_0\right) (r W_0)^{2 U_0-2 K_0}}{r^2}\right)^n\bigg).~~
\end{eqnarray}
If we plot these quantities in Fig. \ref{case3_1} for some values of the constants, we see that although the energy density is always greater than zero for different values of n, the radial and axial pressures, except for $n=1$ for $p_z$ (it is always negative), have initially positive values but quickly become negative. In addition, the azimuthal pressure is always less than zero with respect to $r$.

\begin{figure}[!h]
	\centering
	\subfigure[]{
		\includegraphics[width=.3\textwidth]{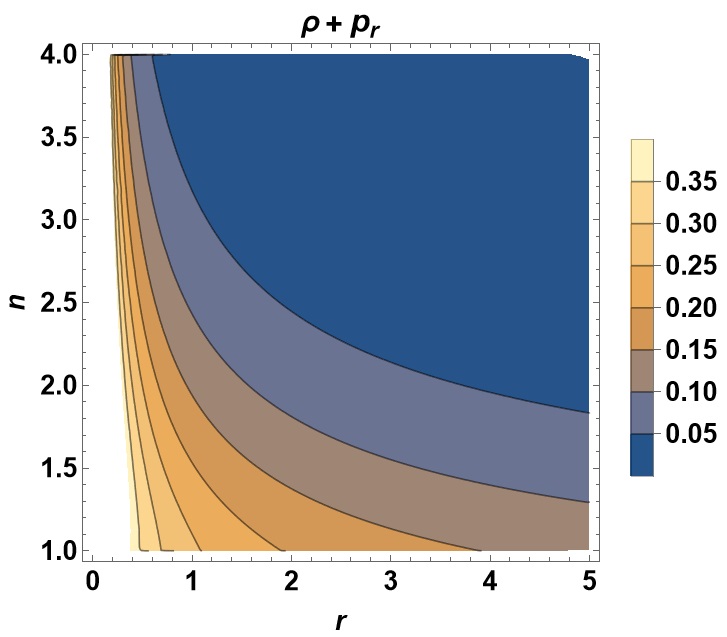}
		\label{necrc3}}
	\subfigure[]{
		\includegraphics[width=.3\textwidth]{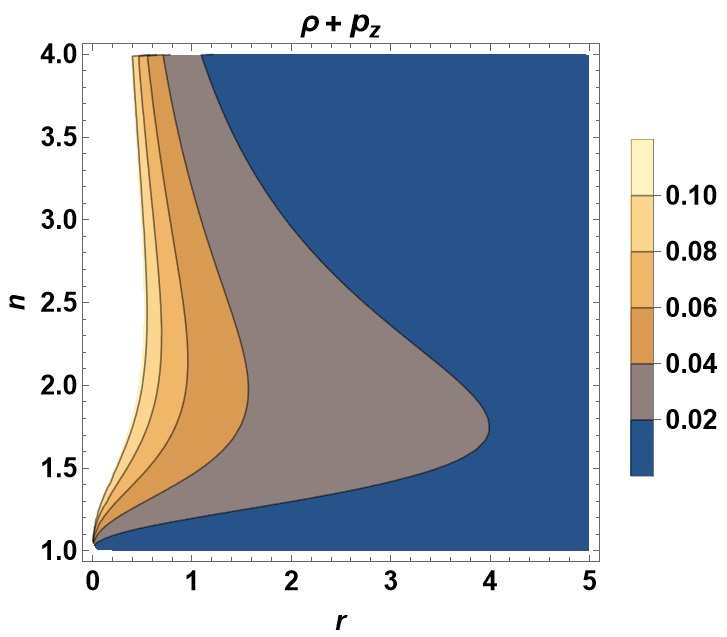}
		\label{neczc3}}
	\subfigure[]{
		\includegraphics[width=.3\textwidth]{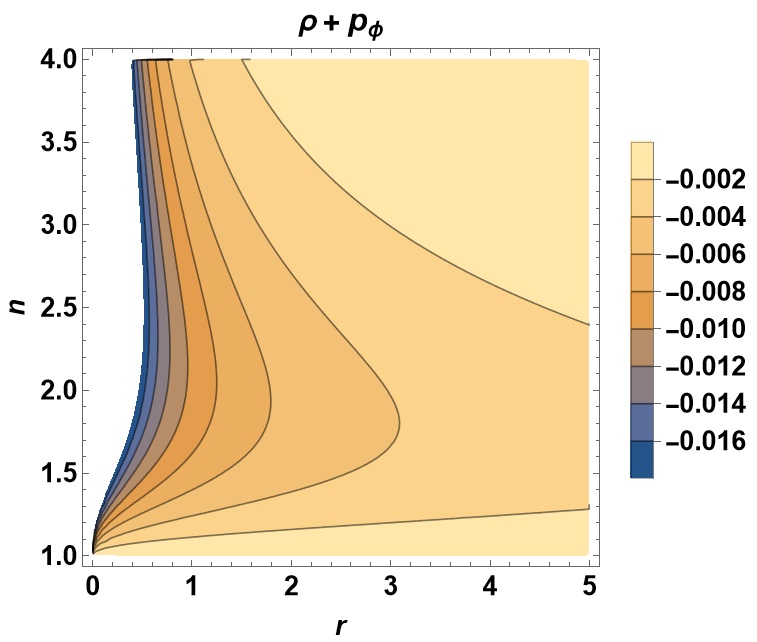}
		\label{necphic3}	}
	\caption{$\rho+p_r$, $\rho+p_z$, $\rho+p_{\phi}$ of Case III of $f=\alpha+\beta Q^{n}$ with $\alpha=-2$, $K_0=-0.4$, $U_0=W_0=0.4$ and $\beta=-1$.}
	\label{case3_2}
\end{figure}

\begin{figure}[!h]
	\centering
	\subfigure[]{
		\includegraphics[width=.3\textwidth]{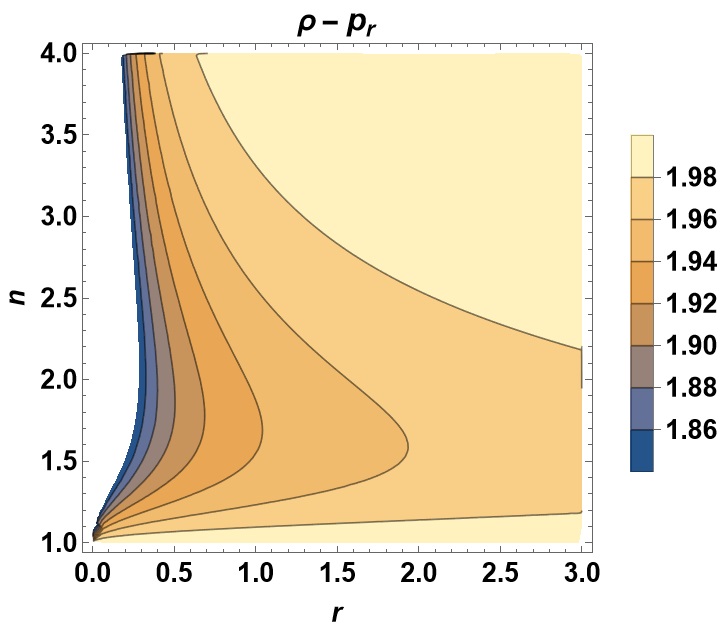}
	}
	\subfigure[]{
		\includegraphics[width=.3\textwidth]{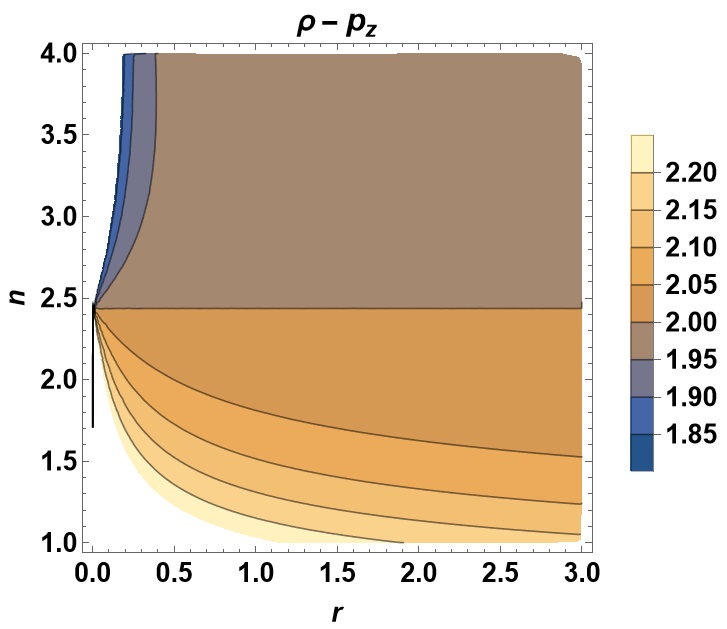}
	}
	\subfigure[]{
		\includegraphics[width=.3\textwidth]{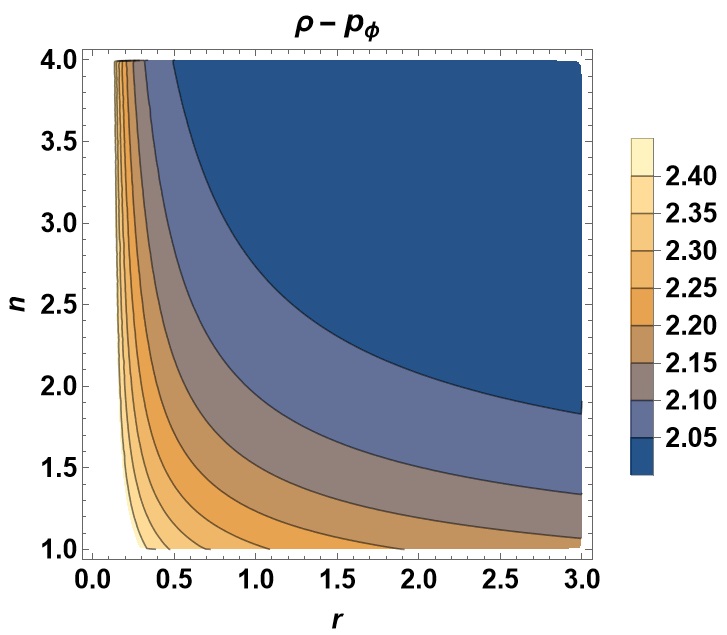}
	}
	\caption{$\rho-p_r$, $\rho-p_z$, $\rho-p_{\phi}$ of Case III of $f=\alpha+\beta Q^{n}$ with $\alpha=-2$, $K_0=-0.4$, $U_0=W_0=0.4$ and $\beta=-1$.}
	\label{case3_3}
\end{figure}

\begin{figure}[!h]
	\centering
	\subfigure[]{
		\includegraphics[width=.3\textwidth]{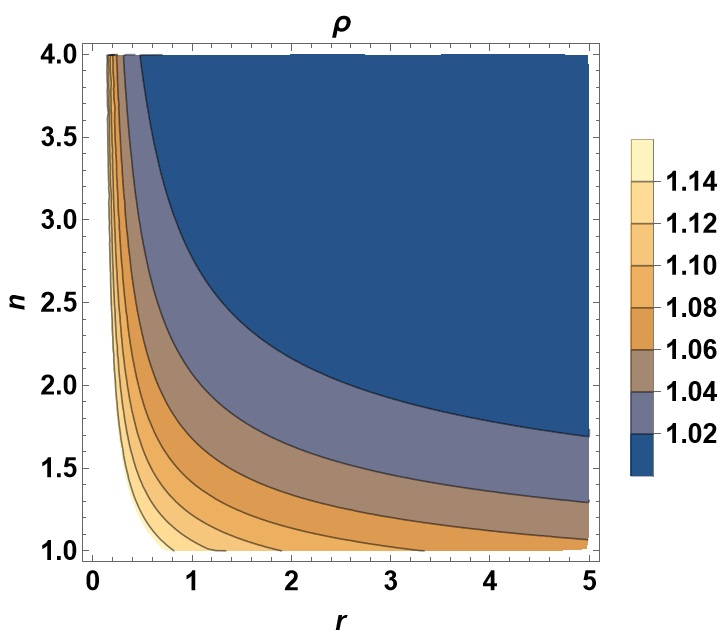}
		\label{wecc3}}
	\subfigure[]{
		\includegraphics[width=.3\textwidth]{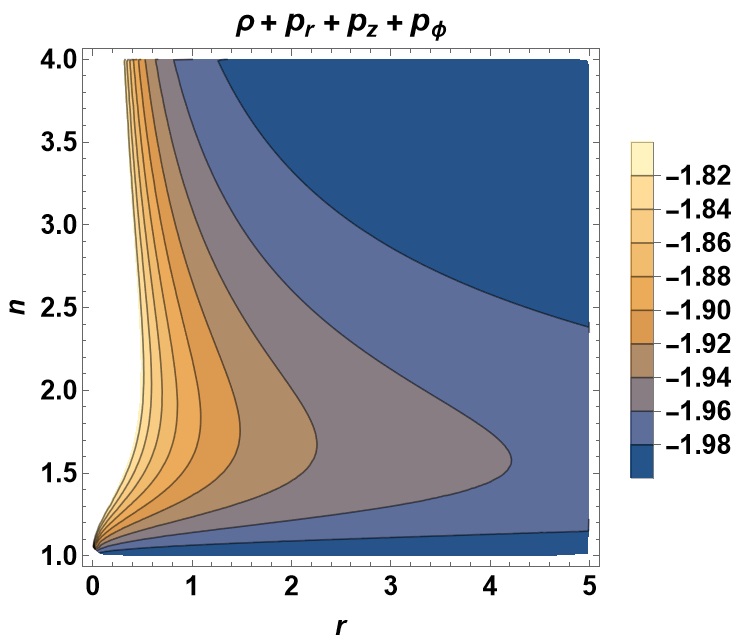}
		\label{secc3}}
	\caption{$\rho$, and $\rho+p_r+p_z+p_{\phi}$ of Case III of $f=\alpha+\beta Q^{n}$ with $\alpha=-2$, $K_0=-0.4$, $U_0=W_0=0.4$ and $\beta=-1$.}
	\label{case3_4}
\end{figure}

$f(Q)$ in the power law solution, the necessary energy condition statements are plotted in Figs. \ref{case3_2}-\ref{case3_4}. Although $\rho+p_r$ and $\rho+p_z$ are always positive, which means that NEC for $r$ and $z$ directions are satisfied in Figs. \ref{necrc3}, \ref{neczc3}, NEC for azimuthal direction is violated because $\rho+p_{\phi}$ is always negative in Fig. \ref{necphic3}. Similarly, WEC is violated for the $\phi$ direction but is satisfied for the radial and axial directions because of $\rho\geq 0$ in Fig. \ref{wecc3}. Despite $\rho-p_n\geq 0$ being satisfied for all directions in Fig. \ref{case3_3}, DEC is violated for the azimuthal direction but is satisfied for other directions. Furthermore, SEC is violated for case III because of the negative nature of $\rho+\sum_{n}p_n$ in Fig. \ref{secc3}. Finally, we can say that NEC, WEC, DEC and SEC are violated for cosmic strings in the $f(Q)=\alpha+\beta Q^n$ solution.

\newpage
\section{$f(Q)=\beta Q e^{\frac{\lambda}{Q}}$ Gravity}\label{section5}
Another function of $f(Q)$ is chosen in the exponential form, and three of the cases mentioned above are discussed for this form. 
\subsection{Case I}
In case I, we use metric functions of (\ref{wc1})-(\ref{uc1}) and the energy density and the directional pressures are given;
\begin{eqnarray}
	\rho&=&-\frac{\beta U_0^2 W_1^2 e^{ \left(\frac{\lambda (r W_0+W_1)^4}{2 U_0 W_1 (2 r W_0 (U_0+W_0)+W_1 (U_0+2 W_0))}\right)}}{2 (r W_0+W_1)^7 (r (U_0+W_0)+W_1)^6 \left(\frac{W_0^2}{r W_0+W_1}-\frac{(U_0+W_0)^2 (r W_0+W_1)}{(r (U_0+W_0)+W_1)^2}\right)^3}\nonumber\\
	&& \Bigg(-2 \lambda^2 W_0 (r W_0+W_1)^8\nonumber\\
	&&\times (r (U_0+W_0)+W_1) (3 r W_0 (U_0+W_0)+W_1 (U_0+3 W_0))\nonumber\\
	&&-U_0 W_1 (2 r W_0 (U_0+W_0)+W_1 (U_0+2 W_0))^2\nonumber\\
	&&\times \bigg(2 U_0 W_1 (2 r W_0 (U_0+W_0)+W_1 (U_0+2 W_0)) (4 r W_0 (U_0+W_0)+W_1 (3 U_0+4 W_0))\nonumber\\
	&&-2 \lambda (U_0+W_0) (r W_0+W_1)^5\bigg)\Bigg) ,\\
	p_r&=& \frac{\beta  e^{ \left(\frac{\lambda (r W_0+W_1)^4}{2 U_0 W_1 (2 r W_0 (U_0+W_0)+W_1 (U_0+2 W_0))}\right)}}{(r W_0+W_1)^4}\bigg(\lambda (r W_0+W_1)^4-U_0 W_1 (2 r W_0 (U_0+W_0)\nonumber\\
	&&+W_1 (U_0+2 W_0))\bigg), \\
	p_z&=&\frac{1}{2} \beta e^{ \left(\frac{\lambda (r W_0+W_1)^4}{2 U_0 W_1 (2 r W_0 (U_0+W_0)+W_1 (U_0+2 W_0))}\right)} \Bigg(\frac{2 U_0 W_1 (2 r W_0 (U_0+W_0)+W_1 (U_0+2 W_0))}{(r W_0+W_1)^4}\nonumber\\
	&&-\frac{\lambda^2 W_0^2 (r W_0+W_1)^4 (r (U_0+W_0)+W_1)^2 (3 r W_0 (U_0+W_0)+W_1 (U_0+3 W_0))}{U_0^2 W_1^2 (2 r W_0 (U_0+W_0)+W_1 (U_0+2 W_0))^3}\Bigg)  , \\
	p_{\phi}&=&\frac{1}{2} \beta e^{ \left(\frac{\lambda (r W_0+W_1)^4}{2 U_0 W_1 (2 r W_0 (U_0+W_0)+W_1 (U_0+2 W_0))}\right)} \Bigg(\frac{2 U_0 W_1 (2 r W_0 (U_0+W_0)+W_1 (U_0+2 W_0))}{(r W_0+W_1)^4}\nonumber\\
	&&-\frac{\lambda^2 W_0^2 (r W_0+W_1)^4 (r (U_0+W_0)+W_1)^2 (3 r W_0 (U_0+W_0)+W_1 (U_0+3 W_0))}{U_0^2 W_1^2 (2 r W_0 (U_0+W_0)+W_1 (U_0+2 W_0))^3}\Bigg) .
\end{eqnarray}
In this case $p_z$ and $p_{\phi}$ are equal, similar to a power series solution. These quantities are plotted for different values of the parameter $\lambda$ to analyse their characteristics. The energy density remains positive for both negative and positive values of the parameter $\lambda$. On the other hand, the radial pressure initially positive then turns to negative values for positive values of $\lambda$ but approaches to zero for $\lambda\leqslant 0$. In addition, the axial and azimuthal pressures have initially negative values, but for $\lambda>0 $ their values increase rapidly in a positive direction, for $\lambda \leqslant 0$ their values tend to approach zero.
\begin{figure}[!h]
	\centering
	\subfigure[]{
		\includegraphics[width=.4\textwidth]{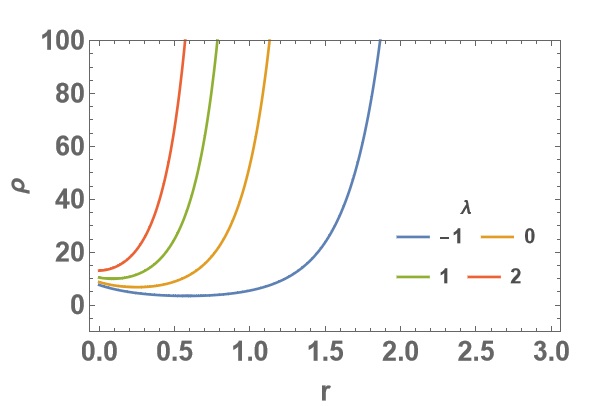}
	}
	\subfigure[]{
		\includegraphics[width=.4\textwidth]{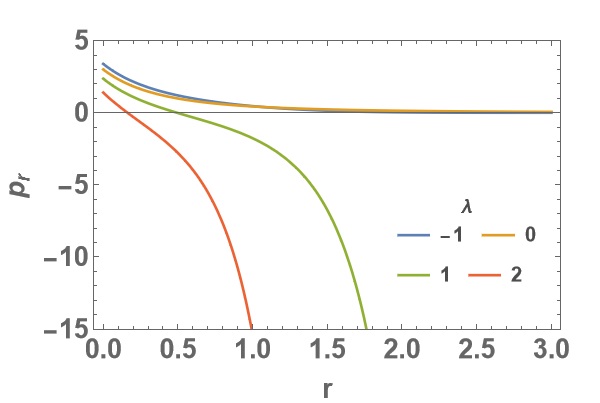}
	}
	
	\subfigure[]{
		\includegraphics[width=.4\textwidth]{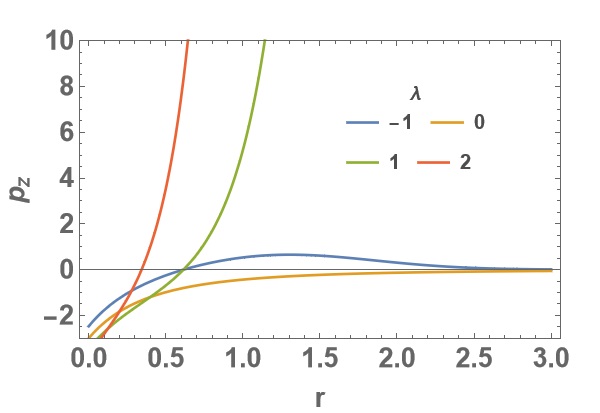}
	}
	\subfigure[]{
		\includegraphics[width=.4\textwidth]{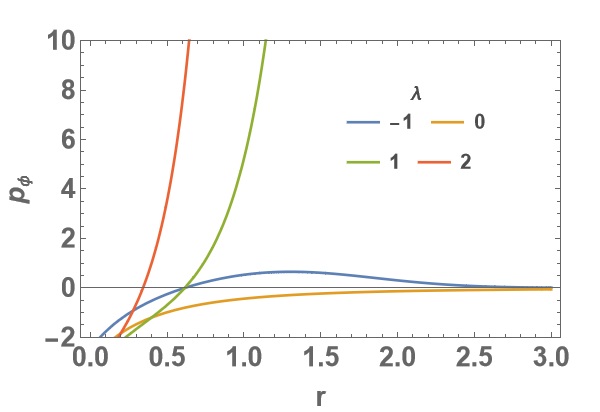}
	}
	\caption{Energy density and radial, axial and azimuthal pressures in Case I for $f(Q)=\beta Q e^{\frac{\lambda}{Q}}$ with $\beta=-1$, $W_0=W_1=U_0=1$.}
	\label{ex_case1_1}
\end{figure} 

\begin{figure}[!h]
	\centering
	\subfigure[]{
		\includegraphics[width=.3\textwidth]{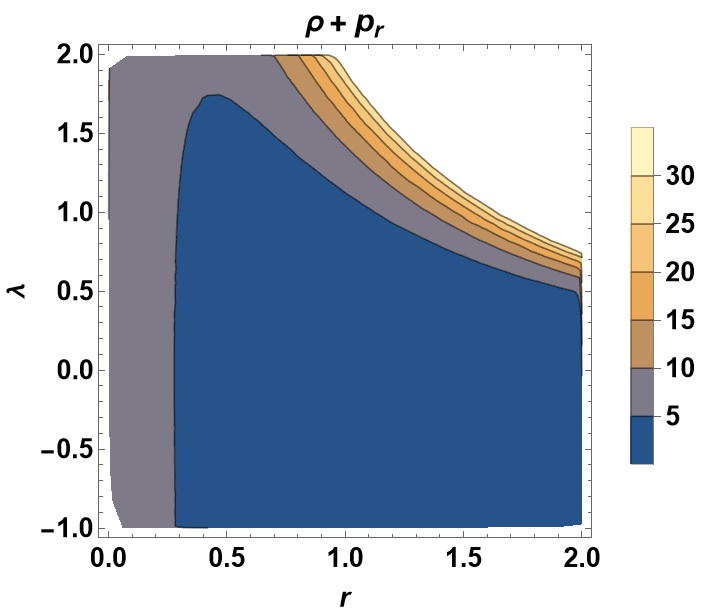}
		\label{necrexc1}}
	\subfigure[]{
		\includegraphics[width=.3\textwidth]{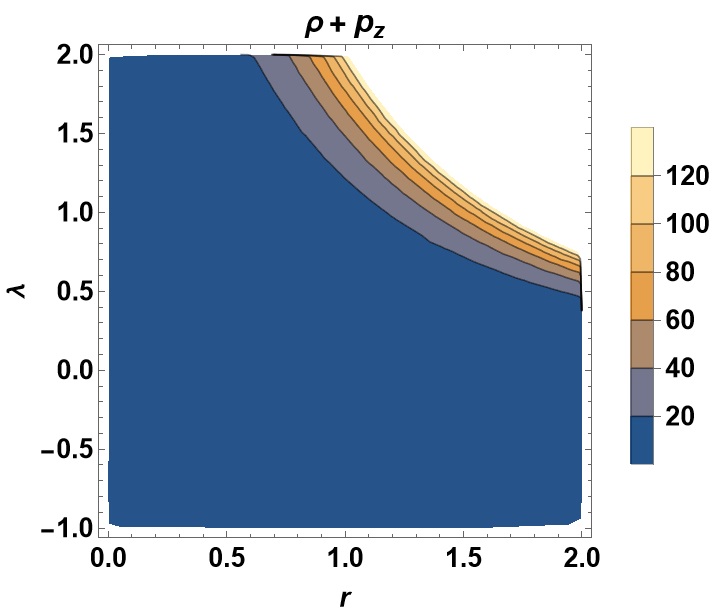}
		\label{neczexc1}}
	\subfigure[]{
		\includegraphics[width=.3\textwidth]{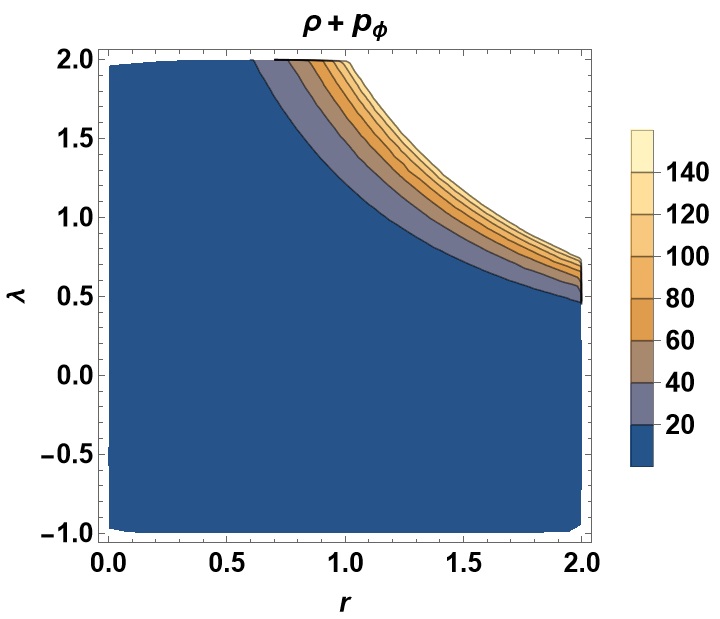}
		\label{necphiexc1}	}
	\caption{Necessary statements of NEC; $\rho+p_r$, $\rho+p_z$, $\rho+p_{\phi}$ of Case I of $f=\beta Q e^{\frac{\lambda}{Q}}$ with $W_0=W_1=U_0=1$ and $\beta=-1$.}
	\label{ex_case1_2}
\end{figure}

\begin{figure}[!h]
	\centering
	\subfigure[]{
		\includegraphics[width=.3\textwidth]{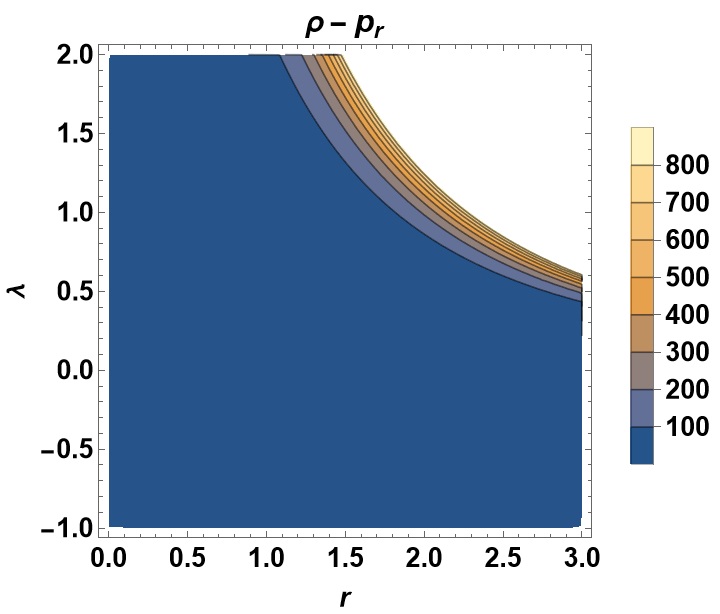}
	}
	\subfigure[]{
		\includegraphics[width=.3\textwidth]{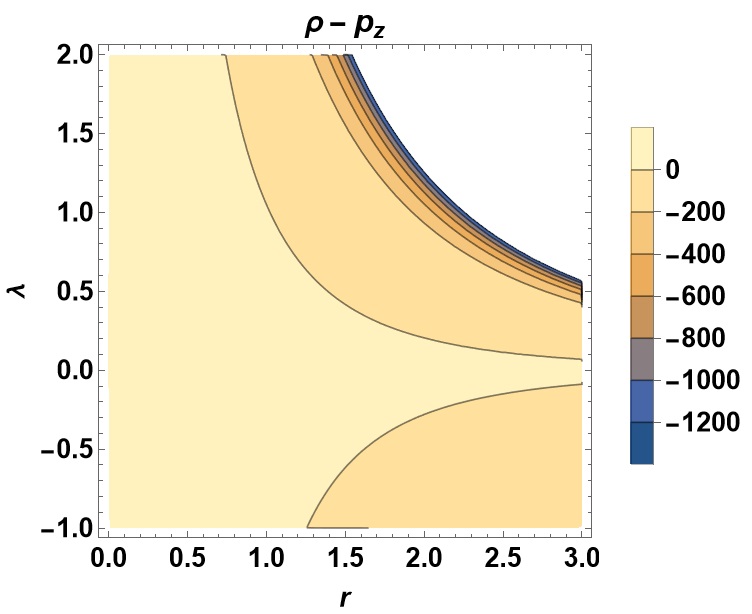}
	}
	\subfigure[]{
		\includegraphics[width=.3\textwidth]{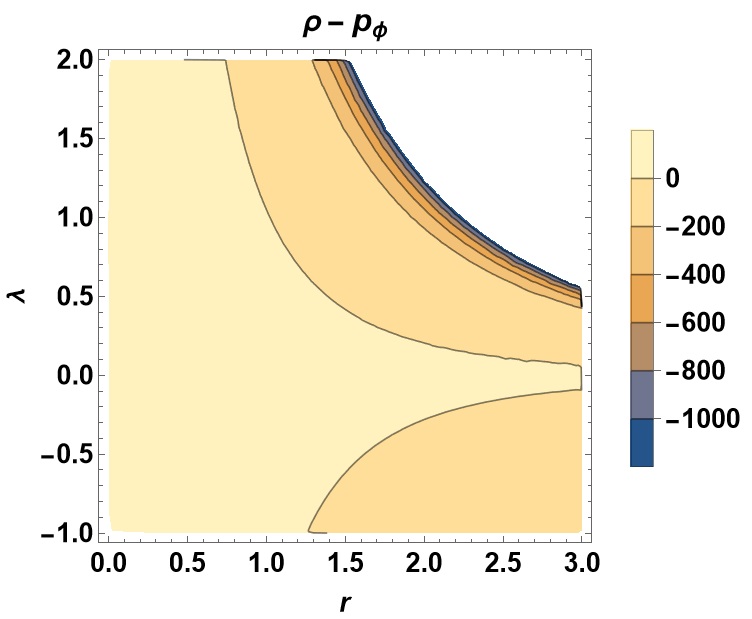}
	}
	\caption{Necessary statements of DEC; $\rho-p_r$, $\rho-p_z$, $\rho-p_{\phi}$  of Case I of $f(Q)=\beta Q e^{\frac{\lambda}{Q}}$ with $W_0=W_1=U_0=1$ and $\beta=-1$.}
	\label{ex_case1_3}
\end{figure}

\begin{figure}[!h]
	\centering
	\subfigure[]{
		\includegraphics[width=.3\textwidth]{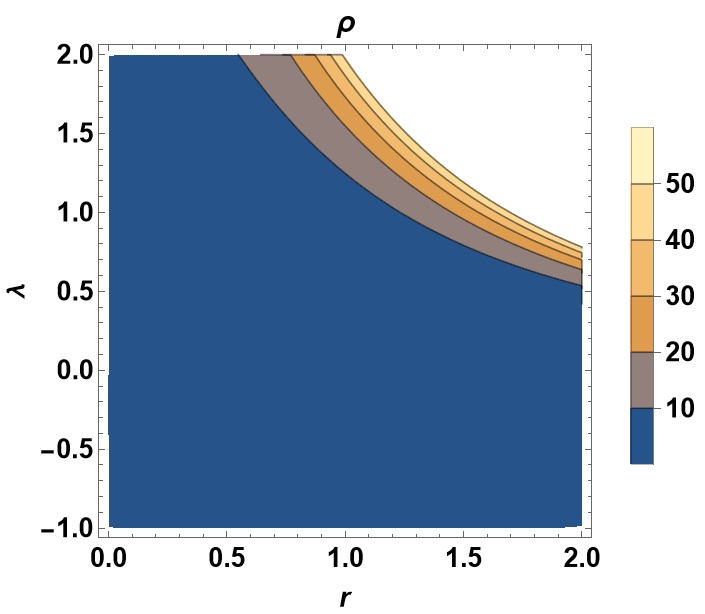}
		\label{ex_wecc1}}
	\subfigure[]{
		\includegraphics[width=.3\textwidth]{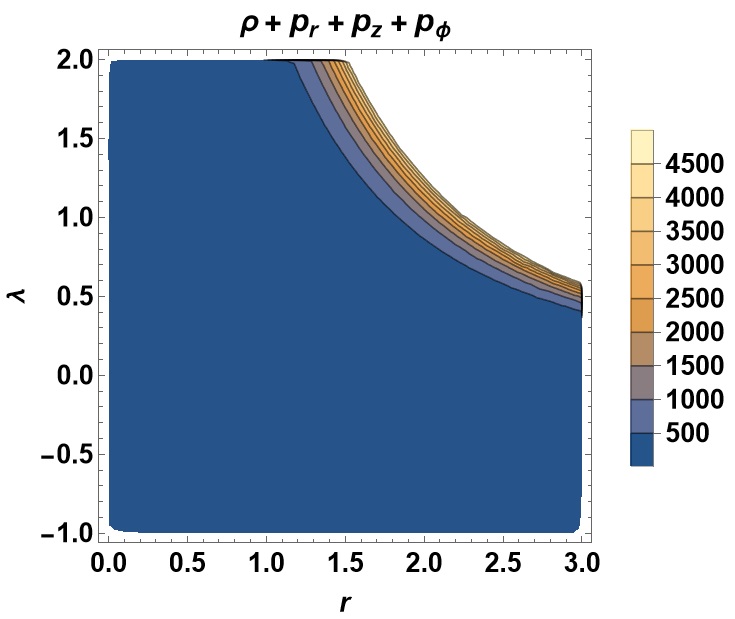}
		\label{ex_secc1}}
	\caption{Necessary statements of WEC and SEC $\rho$, and $\rho+p_r+p_z+p_{\phi}$ of Case I of $f=\beta Q e^{\frac{\lambda}{Q}}$ with $W_0=W_1=U_0=1$ and $\beta=-1$.}
	\label{ex_case1_4}
\end{figure}

The obligatory statements of theenergy conditions are shown in Figs. \ref{ex_case1_2}-\ref{ex_case1_4}. Although, NEC in Fig. \ref{ex_case1_2}, WEC in Fig. \ref{ex_wecc1} and SEC in Fig. \ref{ex_secc1} are satisfied for all directions, DEC is satisfied for the radial direction but is particularly satisfied for the azimuthal and axial directions in Fig. \ref{ex_case1_3}. Therefore, even though NEC, WEC and DEC are satisfied, SEC is violated in this case.

\subsection{Case II}
We introduce the metric potentials of the LC solution eqns. (\ref{wc2}-\ref{uc2}) and obtain the energy density and the directional pressures as;
\begin{eqnarray}
	\rho&=&\frac{\beta r^{-2 (K_0+U_0+1)} e^{-\frac{\lambda r^{2 K_0-2 U_0+2}}{2 K_0-2 U_0^2}}}{2 \left(K_0-U_0^2\right)^2} \bigg( \lambda^2 r^{4 K_0+4} (K_0-U_0+1) (K_0-2 U_0+1)\nonumber\\
	&&+2  \left(K_0-U_0^2\right)^3 r^{4 U_0}\bigg),\\
	p_r&=&\beta r^{-2 (K_0+1)} e^{-\frac{\lambda r^{2 K_0-2 U_0+2}}{2 K_0-2 U_0^2}} \left(\lambda r^{2 K_0+2}+\left(K_0-U_0^2\right) r^{2 U_0}\right),\\
	p_z&=&-\frac{\beta r^{-2 (K_0+U_0+1)} e^{-\frac{\lambda r^{2 K_0-2 U_0+2}}{2 K_0-2 U_0^2}} \left(\lambda^2 r^{4 K_0+4} (K_0-U_0+1)+2 \left(K_0-U_0^2\right)^3 r^{4 U_0}\right)}{2 \left(K_0-U_0^2\right)^2},\\
	p_{\phi}&=&\frac{\beta r^{-2 (K_0+U_0+1)} e^{-\frac{\lambda r^{2 K_0-2 U_0+2}}{2 K_0-2 U_0^2}} \left(2  \left(U_0^2-K_0\right)^3 r^{4 U_0}- K_0 \lambda^2 r^{4 K_0+4} (K_0-U_0+1)\right)}{2 \left(K_0-U_0^2\right)^2}.
\end{eqnarray}

\begin{figure}[!h]
	\centering
	\subfigure[]{
		\includegraphics[width=.4\textwidth]{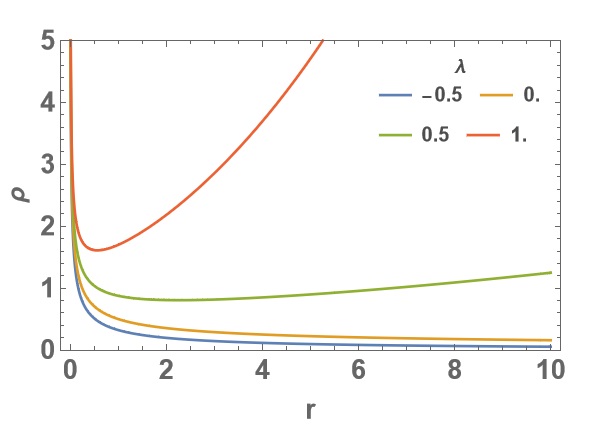}
	}
	\subfigure[]{
		\includegraphics[width=.4\textwidth]{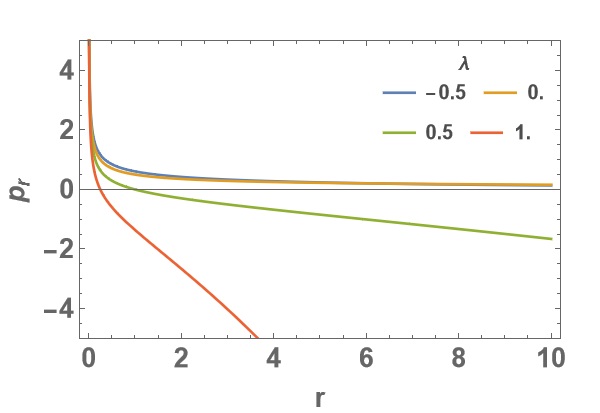}
	}
	
	\subfigure[]{
		\includegraphics[width=.4\textwidth]{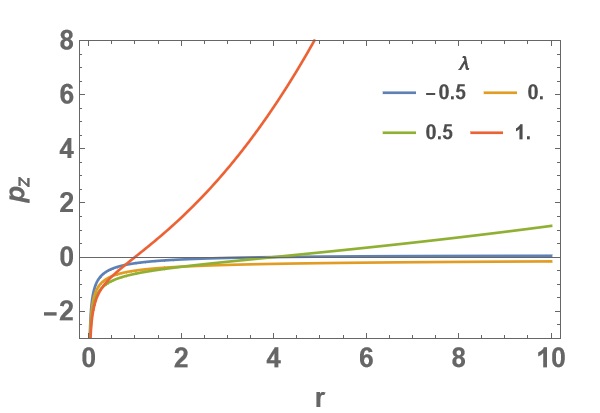}
	}
	\subfigure[]{
		\includegraphics[width=.4\textwidth]{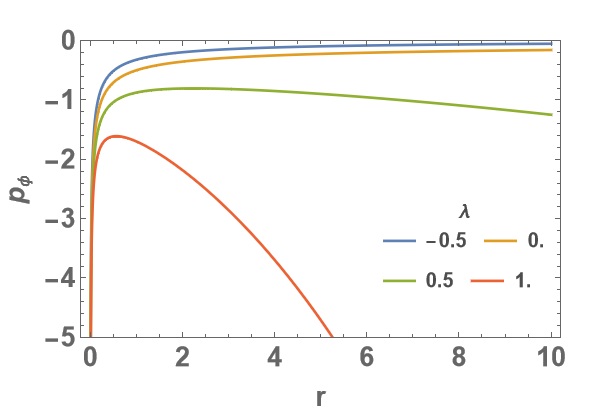}
	}
	\caption{Energy density and radial, axial and azimuthal pressures in Case II for $f(Q)=\beta Q e^{\frac{\lambda}{Q}}$ with $\beta=-1$, $K_0=-0.25$ and $U_0=0.5$.}
	\label{ex_case2_1}
\end{figure}

We plot these quantities for positive and negative values of $\lambda$ in Fig. \ref{ex_case2_1}. Despite the energy density always having positive values, the radial pressure of all $\lambda$ has positive values for small $r$, and $\lambda\leq 0$ approaches zero, but $\lambda>0$ tends to negative as $r$ increases.  Furthermore, the azimuthal pressure always has negative values for all $\lambda$, but the axial pressure initially has negative values for all $\lambda$, then for $\lambda>0$ it becomes positive, otherwise it approaches zero. 

\begin{figure}[!h]
	\centering
	\subfigure[]{
		\includegraphics[width=.3\textwidth]{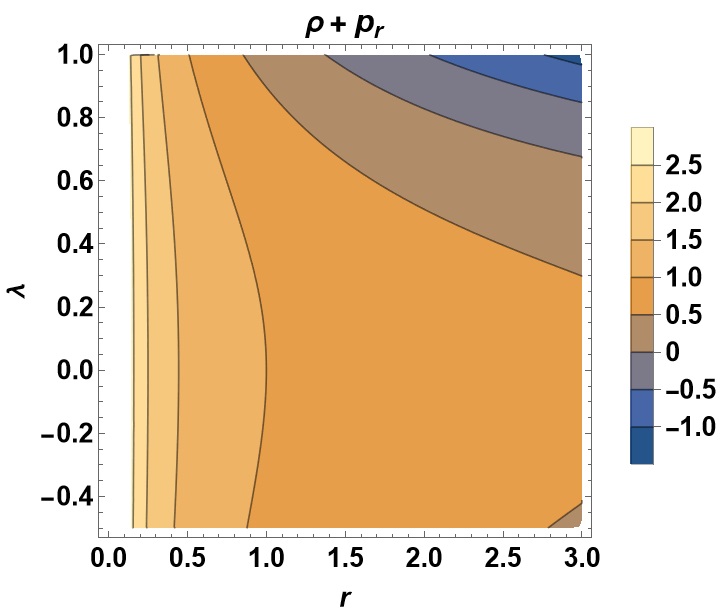}
		\label{necrexc2}}
	\subfigure[]{
		\includegraphics[width=.3\textwidth]{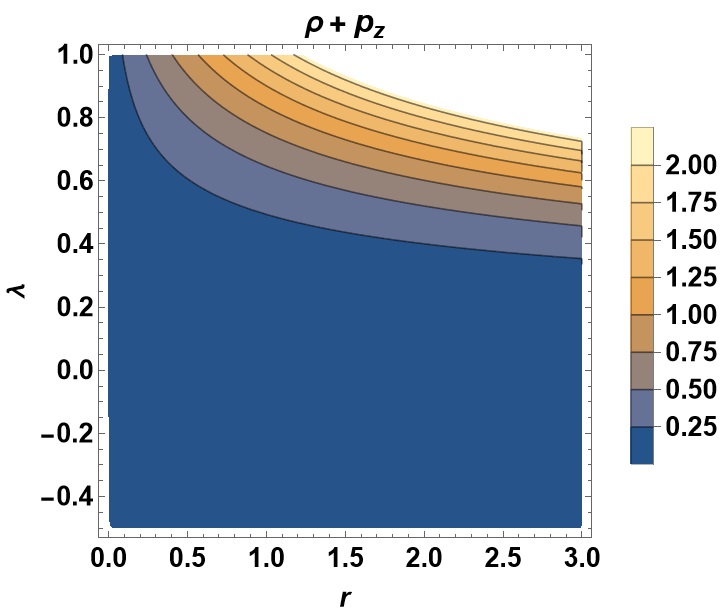}
		\label{neczexc2}}
	\caption{Necessary statements of NEC; $\rho+p_r$, $\rho+p_z$, $\rho+p_{\phi}$ of Case II of $f=\beta Q e^{\frac{\lambda}{Q}}$ with $\beta=-1$, $K_0=-0.25$ and $U_0=0.5$.}
	\label{ex_case2_2}
\end{figure}

\begin{figure}[!h]
	\centering
	\subfigure[]{
		\includegraphics[width=.3\textwidth]{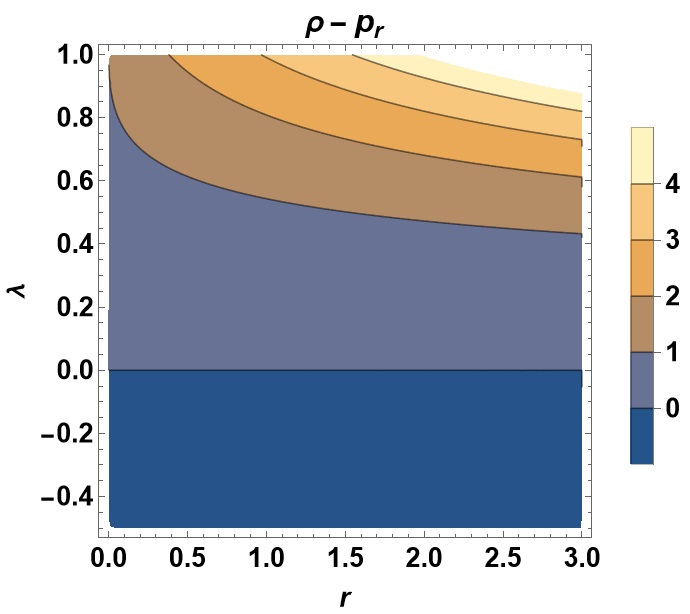}
	}
	\subfigure[]{
		\includegraphics[width=.3\textwidth]{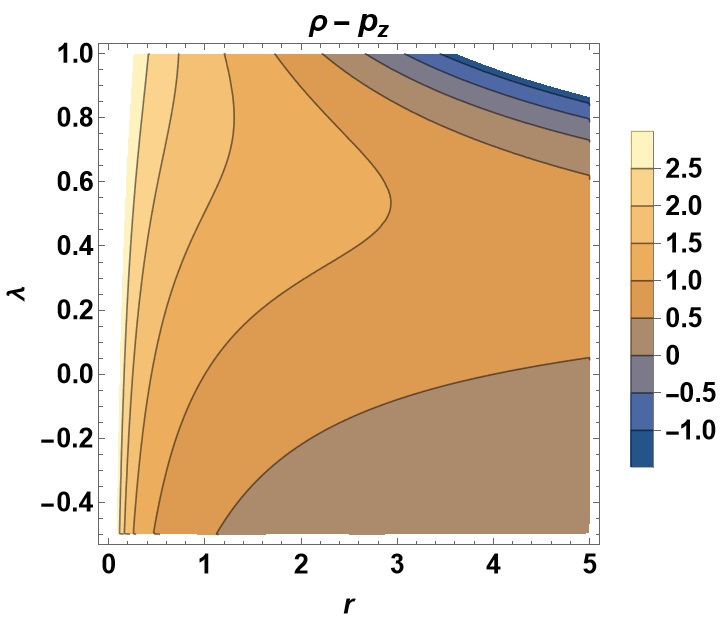}
	}
	\subfigure[]{
		\includegraphics[width=.3\textwidth]{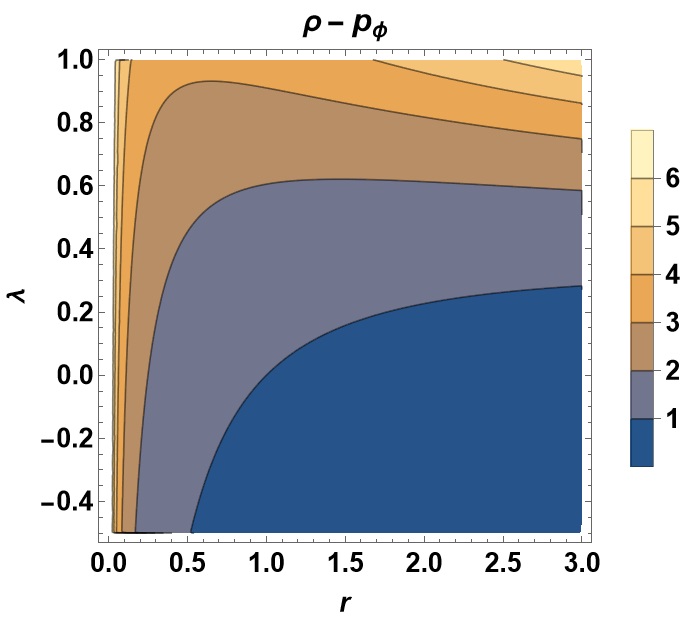}
	}
	\caption{Necessary statements of DEC; $\rho-p_r$, $\rho-p_z$, $\rho-p_{\phi}$  of Case II of $f(Q)=\beta Q e^{\frac{\lambda}{Q}}$ with $\beta=-1$, $K_0=-0.25$ and $U_0=0.5$.}
	\label{ex_case2_3}
\end{figure}

\begin{figure}[!h]
	\centering
	\subfigure[]{
		\includegraphics[width=.3\textwidth]{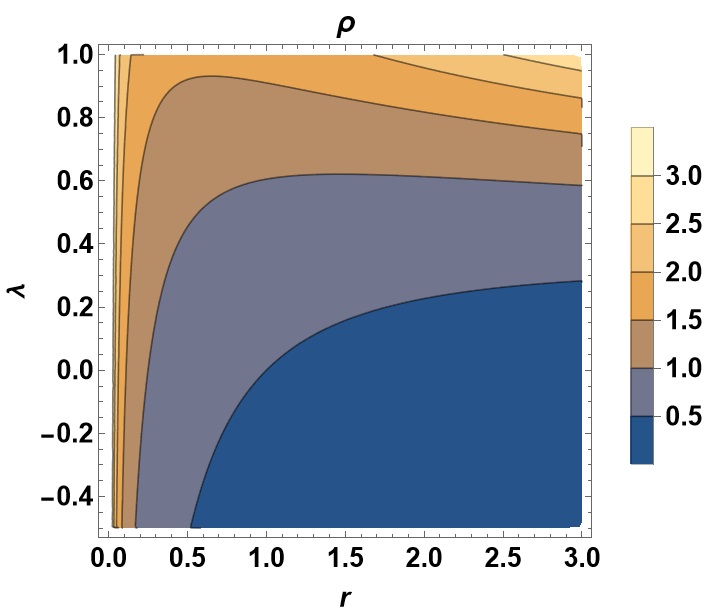}
		\label{ex_wecc2}}
	\subfigure[]{
		\includegraphics[width=.3\textwidth]{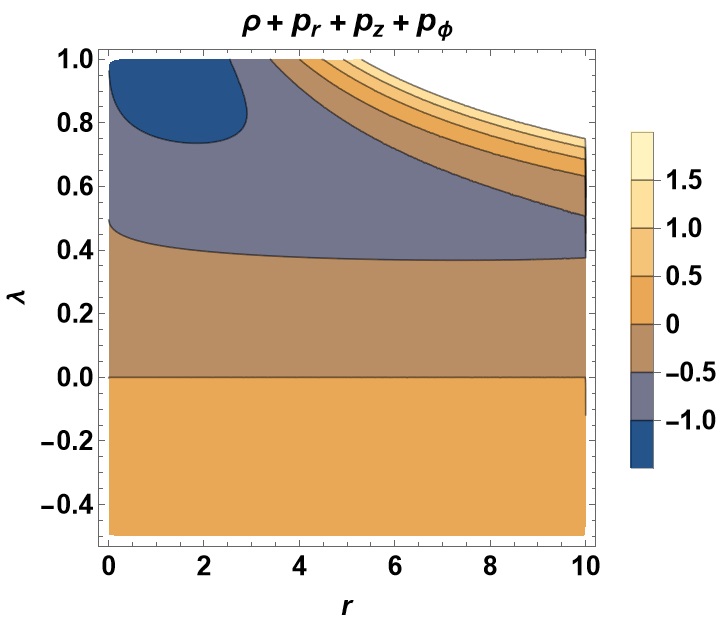}
		\label{ex_secc2}}
	\caption{Necessary statements of WEC and SEC $\rho$, and $\rho+p_r+p_z+p_{\phi}$ of Case II of $f=\beta Q e^{\frac{\lambda}{Q}}$ with $\beta=-1$, $K_0=-0.25$ and $U_0=0.5$.}
	\label{ex_case2_4}
\end{figure}

The obligatory statements of the energy conditions are shown in the figures \ref{ex_case2_2}-\ref{ex_case2_4}.  The NEC for the azimuthal direction is satisfied for given values of the constants, as $\rho+p_{\phi}=0$. In addition, the NEC for the $z$-direction is satisfied, but the NEC for the radial direction is violated for large $r$ and large $\lambda$ values in Figs. \ref{neczexc2} and \ref{necrexc2}, respectively. While the positive nature of the energy density in Figure \ref{ex_wecc2}, WEC is satisfied in the $z, \phi$ directions, it is particularly satisfied for the $r$ direction. 
Similarly, DEC is satisfied for azimuthal direction, but it is especially violated
radial and axial directions where necessary statements are plotted in Fig. \ref{ex_case2_3}. Interestingly, the $\rho+\sum_{n}p_n \geq 0$ statement of SEC is satisfied for all $r$ values as $\lambda\leq 0$, but it is violated especially for $\lambda\geq 0$ in Fig. \ref{ex_secc2}. Finally, in the LC solution of the exponential $f(Q)$ gravity, all energy conditions are satisfied in small regions.

\subsection{Case III}
Cosmic strings are discussed for $f(Q)$ has an exponential with the metric potentials of eqns. \ref{wc3}-\ref{uc3}. Three different directional pressures and energy density are yield as;
\begin{eqnarray}
	\rho&=&\frac{\beta r^{-2 (K_0+U_0)-2} W_0^{-2 (K_0+U_0)}e^{-\frac{\lambda r^{2 K_0-2 U_0+2} W_0^{2 K_0-2 U_0}}{2 K_0-2 U_0^2}}}{2 \left(K_0-U_0^2\right)^2}\nonumber\\
	&&\times \bigg( \lambda^2 r^{4 K_0+4} (K_0-U_0+1) (K_0-2 U_0+1) W_0^{4 K_0}+2 \left(K_0-U_0^2\right)^3 r^{4 U_0} W_0^{4 U_0}\bigg),\\
	p_r&=&\beta r^{-2 K_0-2} W_0^{-2 K_0} e^{-\frac{\lambda r^{2 K_0-2 U_0+2} W_0^{2 K_0-2 U_0}}{2 K_0-2 U_0^2}}\nonumber\\
	&&\times \left(\lambda r^{2 K_0+2} W_0^{2 K_0}+\left(K_0-U_0^2\right) r^{2 U_0} W_0^{2 U_0}\right)  ,\\
	p_z&=&-\frac{\beta r^{-2 (K_0+U_0)-2} W_0^{-2 (K_0+U_0)} e^{-\frac{\lambda r^{2 K_0-2 U_0+2} W_0^{2 K_0-2 U_0}}{2 K_0-2 U_0^2}}}{2 \left(K_0-U_0^2\right)^2}\nonumber\\
	&& \bigg(\lambda^2 r^{4 K_0+4} (K_0-U_0+1) W_0^{4 K_0}+2 \left(K_0-U_0^2\right)^3 r^{4 U_0} W_0^{4 U_0}\bigg),\\
	p_{\phi}&=&  -\frac{\beta r^{-2 (K_0+U_0)-2} W_0^{-2 (K_0+U_0)} e^{-\frac{\lambda r^{2 K_0-2 U_0+2} W_0^{2 K_0-2 U_0}}{2 K_0-2 U_0^2}}}{2 \left(K_0-U_0^2\right)^2} \nonumber\\
	&&\bigg(\lambda^2 K_0 r^{4 K_0+4} (K_0-U_0+1) W_0^{4 K_0}+2 \left(K_0-U_0^2\right)^3 r^{4 U_0} W_0^{4 U_0}\bigg).
\end{eqnarray}

\begin{figure}[!h]
	\centering
	\subfigure[]{
		\includegraphics[width=.4\textwidth]{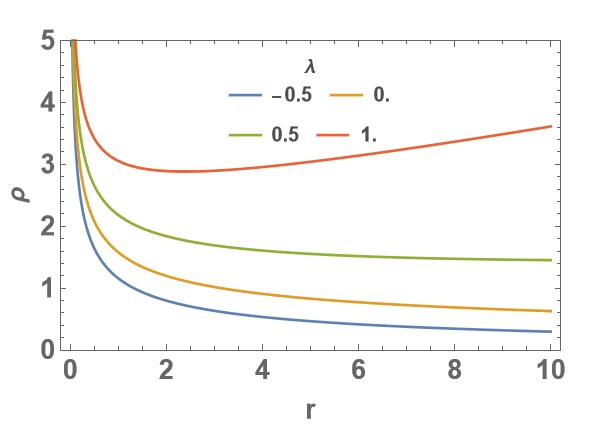}
	}
	\subfigure[]{
		\includegraphics[width=.4\textwidth]{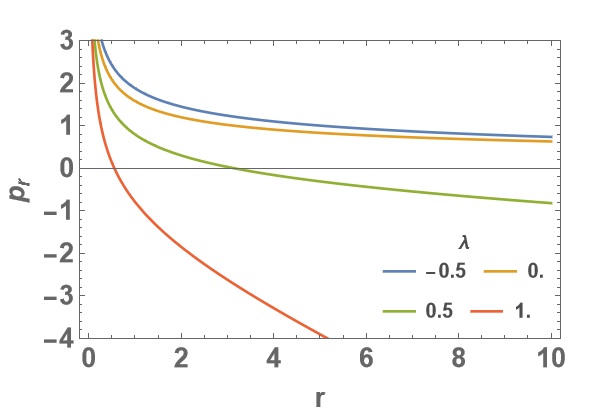}
	}
	
	\subfigure[]{
		\includegraphics[width=.4\textwidth]{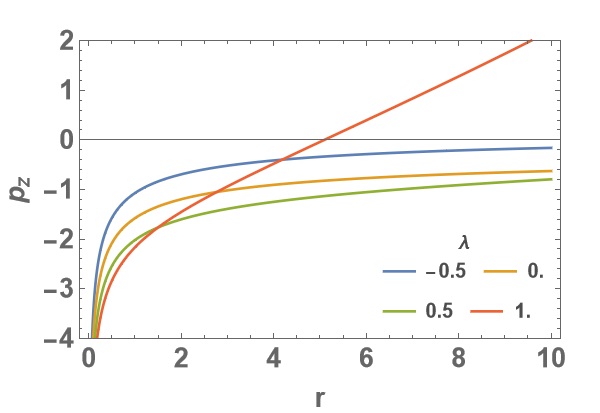}
	}
	\subfigure[]{
		\includegraphics[width=.4\textwidth]{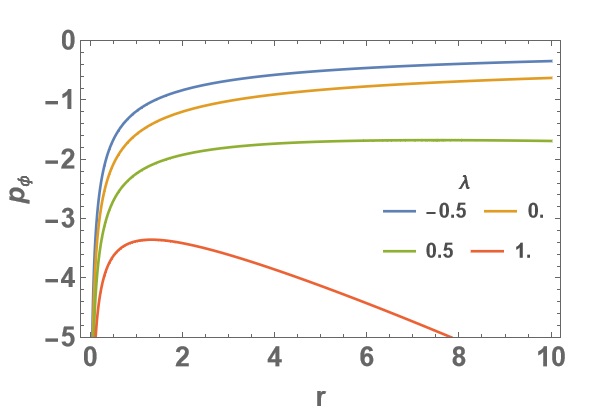}
	}
	\caption{Energy density and radial, axial and azimuthal pressures in Case III for $f(Q)=\beta Q e^{\frac{\lambda}{Q}}$ with $\beta=-2$, $K_0=-0.5$, $W_0=1.2$ and $U_0=0.3$.}
	\label{ex_case3_1}
\end{figure}

We have plotted the energy density and pressures for a few values of $\lambda$ to analyse the cosmic strings in the exponential form of the $f(Q)$ gravity function in Fig. \ref{ex_case3_1}. In addition to $\rho$ remaining positive, the azimuthal pressure remains negative for all values of $\lambda$. The radial pressure is initially positive and becomes negative for $\lambda>0$, but tends to approach zero for $\lambda\leq0$ for large values of $r$. Conversely, the axial pressure is initially negative for small $r$ and then becomes positive for $\lambda>0$ or remains negative for $\lambda\leq0$ as $r$ increases.

\begin{figure}[!h]
	\centering
	\subfigure[]{
		\includegraphics[width=.3\textwidth]{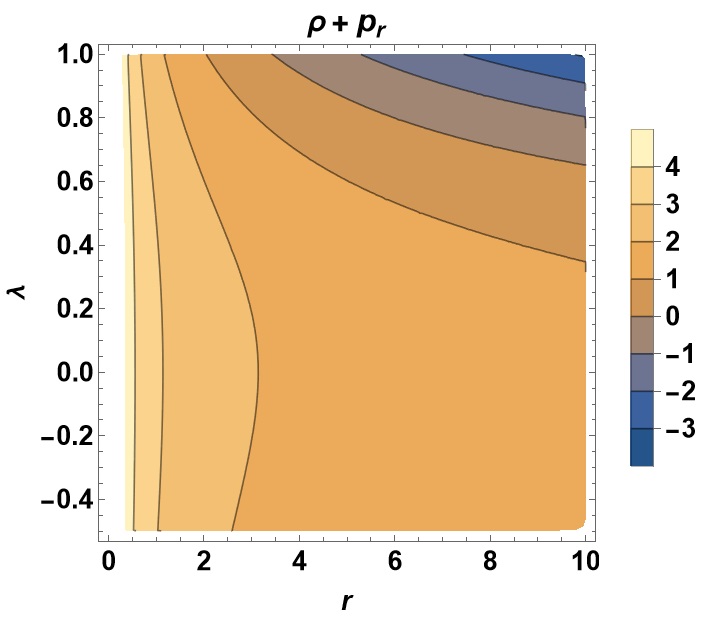}
		\label{necrexc3}}
	\subfigure[]{
		\includegraphics[width=.3\textwidth]{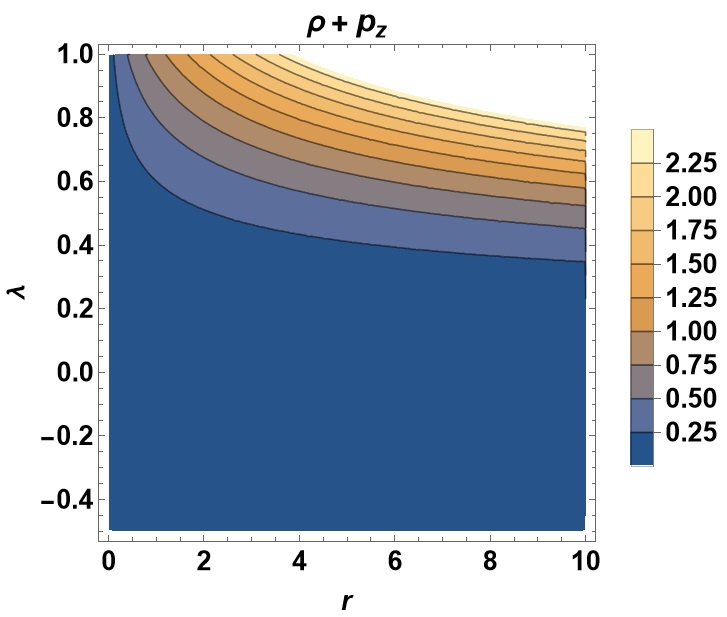}
		\label{neczexc3}}
	\subfigure[]{
		\includegraphics[width=.3\textwidth]{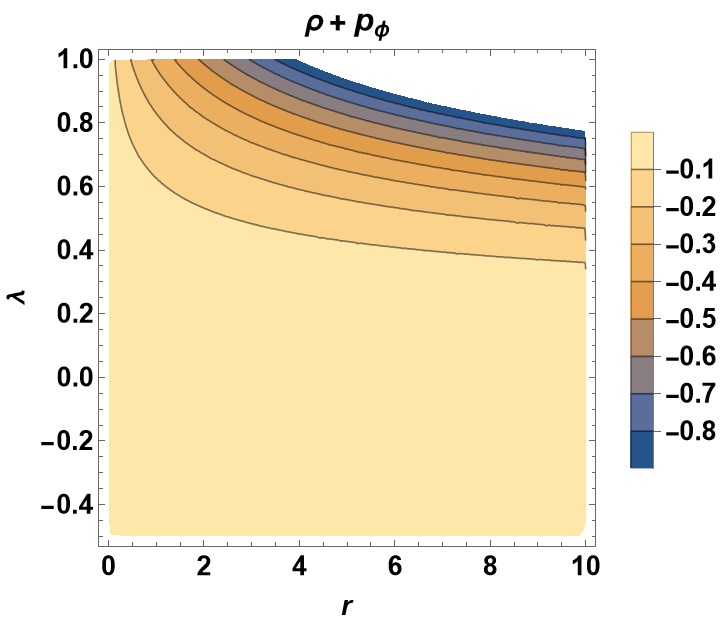}
		\label{necphiexc3}	}
	\caption{Necessary statements of NEC; $\rho+p_r$, $\rho+p_z$, $\rho+p_{\phi}$ of Case III of $f=\beta Q e^{\frac{\lambda}{Q}}$ $\beta=-2$, $K_0=-0.5$, $W_0=1.2$ and $U_0=0.3$.}
	\label{ex_case3_2}
\end{figure}

\begin{figure}[!h]
	\centering
	\subfigure[]{
		\includegraphics[width=.3\textwidth]{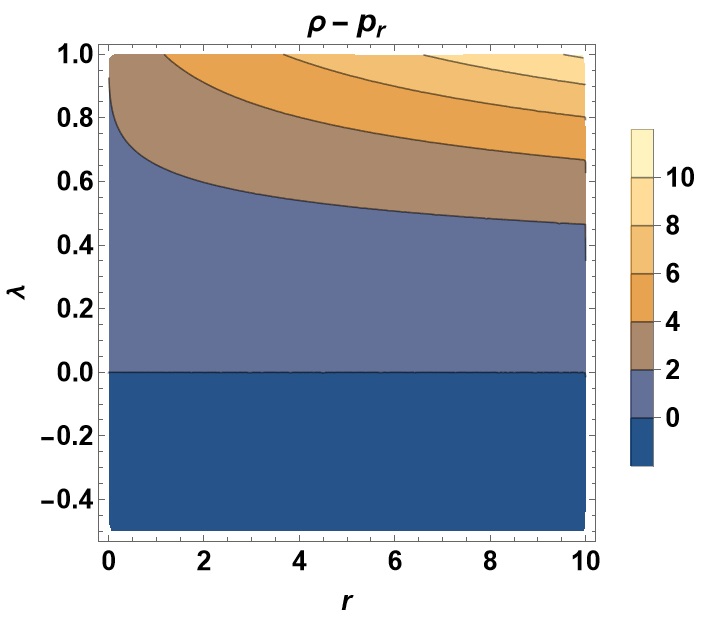}
		\label{ex_decrc3}	}
	\subfigure[]{
		\includegraphics[width=.3\textwidth]{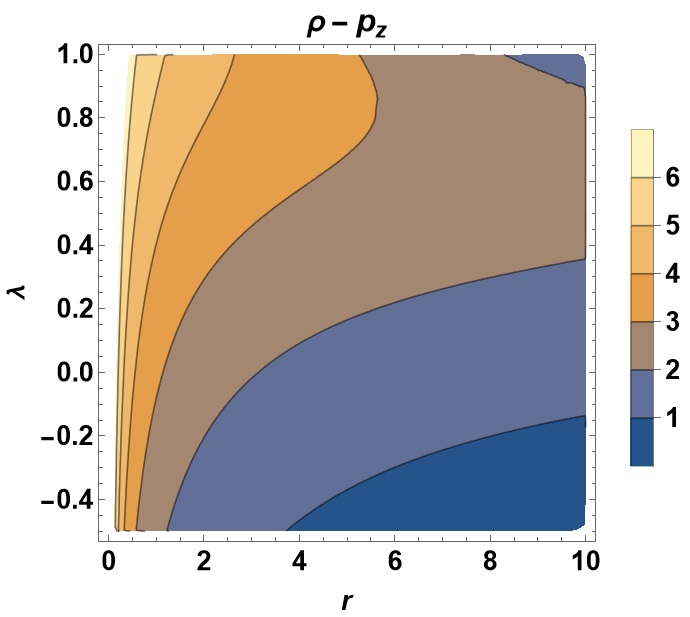}
		\label{ex_deczc3}}
	\subfigure[]{
		\includegraphics[width=.3\textwidth]{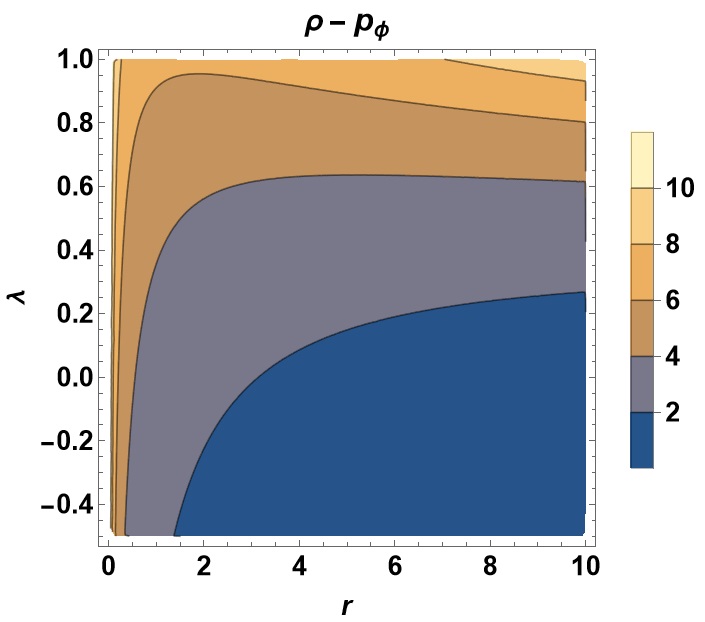}
		\label{ex_decphic3}}
	\caption{Necessary statements of DEC; $\rho-p_r$, $\rho-p_z$, $\rho-p_{\phi}$  of Case III of $f(Q)=\beta Q e^{\frac{\lambda}{Q}}$ with $\beta=-2$, $K_0=-0.5$, $W_0=1.2$ and $U_0=0.3$.}
	\label{ex_case3_3}
\end{figure}

\begin{figure}[!h]
	\centering
	\subfigure[]{
		\includegraphics[width=.3\textwidth]{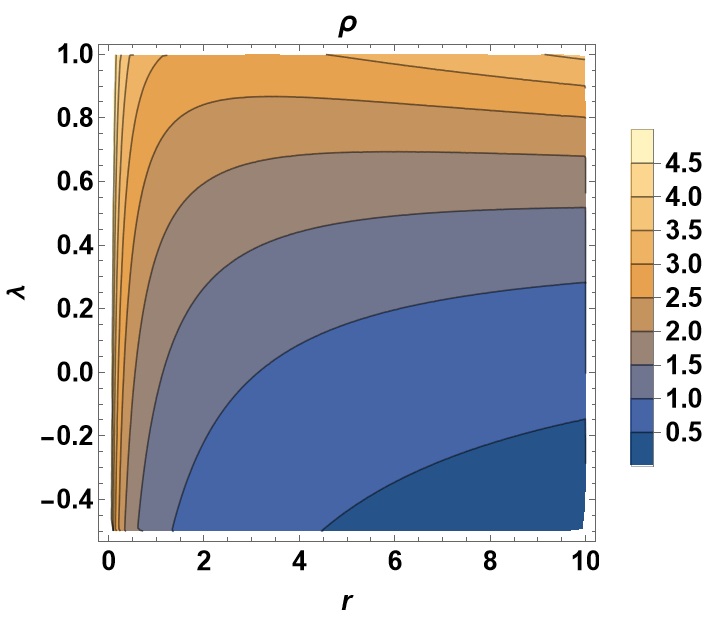}
		\label{ex_wecc3}}
	\subfigure[]{
		\includegraphics[width=.3\textwidth]{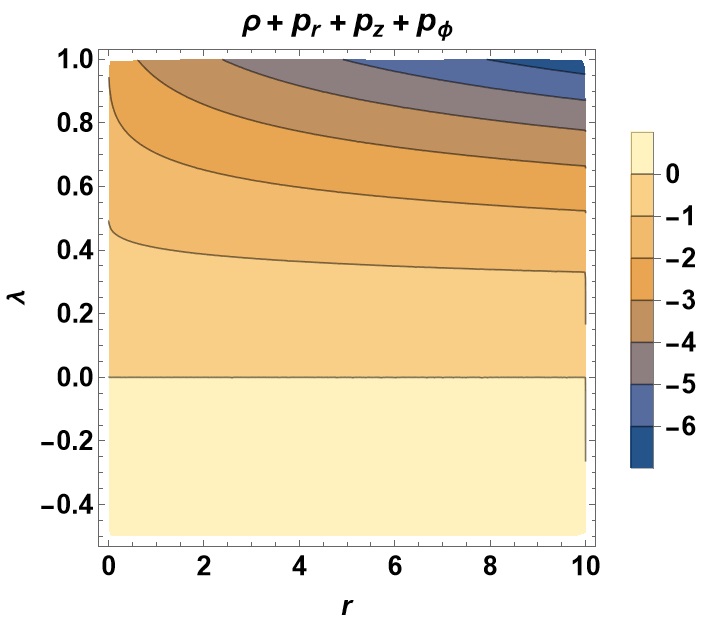}
		\label{ex_secc3}}
	\caption{Necessary statements of WEC and SEC $\rho$, and $\rho+p_r+p_z+p_{\phi}$ of Case III of $f=\beta Q e^{\frac{\lambda}{Q}}$ with$\beta=-2$, $K_0=-0.5$, $W_0=1.2$ and $U_0=0.3$.}
	\label{ex_case3_4}
\end{figure}

Although, the inequalities $\rho-p_{\phi}\geq 0$ and $\rho\geq 0$ are satisfied in Figs. \ref{ex_decphic3} and \ref{ex_wecc3}, the azimuthal directional NEC, WEC and DEC are violated because of the condition $\rho+p_{\phi}\geq 0$ is not satisfied which is shown in Fig. \ref{necphiexc3}, for cosmic strings in exponential $f(Q)$ gravity.On the other hand, NEC in Fig. \ref{neczexc3}, WEC is fully satisfied, but DEC in Fig. \ref{ex_deczc3} is particularly satisfied, for the axial direction. Curiously, NEC and WEC in radial direction are violated in the region of large $\lambda$ and $r$ in Fig. \ref{necrexc3} and the obligatory statement of DEC $\rho-p_r\geq0$ is violated for $\lambda\leq0$ in Fig. \ref{ex_decrc3}. Necessary statement of SEC $\rho+\sum_{n}p_n\geq0$ is satisfied only for $\lambda\leq0$ in Fig. \ref{ex_secc3} that gives radial and axial directional SEC is satisfied for $\lambda\leq0$, but azimuthal direction is violated. Because of these results, we can conclude that cosmic string solution in exponential $f(Q)$ gravity is violated all energy conditions.

\newpage
\section{Conclusion}\label{conc}
In the context of the coincident gauge of $f(Q)$ theory, the static, cylindrically symmetric spacetime has been studied. Interestingly, unlike the field equations of spherically symmetric spacetime in $f(Q)$ theory, this spacetime contains no restriction on the determination of the function of $f(Q)$ or $Q$. Because of this result, the construction of the $f(Q)$ function could take many different forms. For this reason we choose two basic different forms of the function $f(Q)$ to understand the behaviour of this spacetime, as, the power law form $f(Q)=\alpha+\beta Q^n$ and the exponential form $f(Q)=\beta Q e^{\frac{\lambda}{Q}}$. In addition, we introduced the energy-momentum tensor as a perfect fluid with $\rho$ energy density and $p_n$ directional pressures. Three different metric potentials were discussed for both functions of $f(Q)$ and positive energy densities, which is a necessity for regular matter, are achieved by assigning appropriate values of constants. Our results can be summarised as follows;
\begin{itemize}
	\item In Case I, metric potentials such as (\ref{wc1}-\ref{uc1}) were chosen, where axial and azimuthal pressures are equal, and constants for positive energy density were determined in both $f(Q)$ functions. Although the directional pressures have negative values, indicating dark matter and dark energy (or exotic matter) for increasing $r$ in the power-law form of $f(Q)$, they have negative and positive values in the exponential form of $f(Q)$. Furthermore, NEC and WEC are satisfied for both forms of $f(Q)$, DEC is only satisfied in the exponential form. SEC is violated in case I for both solutions. 
	\item In Case II, the metric coefficients were assigned as eqn. (\ref{wc2}-\ref{uc2}), which is a well-known static, cylindrically symmetric solution named Levi-Civita. While the constants were set for positive energy densities, the $p_r, p_z, p_{\phi}$ pressures were negative at large $r$ in the power-law solution. On the other hand, the azimuthal pressure is always negative, while the values of the radial and axial pressures are both negative and positive for different values of $\lambda$ in the exponential solution. Therefore, this case does not contain a regular matter solution. Also, all energy conditions were violated in the power-law form of $f(Q)$, but they were satisfied in small regions in the exponential form.
	\item In Case III, the cosmic string solution whose metric potentials are given in Eqs. (\ref{wc3}-\ref{uc3}) was analysed in the coincident $f(Q)$ theory. When we assigned to obtain positive energy densities, the directional pressures of the power-law solution became negative with increasing $r$. While $p_r, p_z$ have positive values in some regions, $p_{\phi}$ is always negative in the exponential form of $f(Q)$. Thus, the cosmic string solution in the coincident $f(Q)$ theory points to exotic matter. Furthermore, all energy conditions for the cosmic string solution are violated.
\end{itemize}
In contrast to previous studies of cylindrically symmetric spacetime in different theories, we have plotted the energy densities, the directional pressures and the obligatory statements of energy conditions in detailed. The results we found are important in that they not only satisfy the curiosity about the mathematical limits of $f(Q)$ theory, but also show that physically there are mostly no regular matter solutions in this spacetime for this theory.

	\bibliographystyle{ieeetr}
	\bibliography{refs}

\end{document}